\documentclass[a4paper,12pt]{article}

\usepackage{graphicx}
\usepackage{amsmath,amsfonts}
\usepackage{mathtools}
\usepackage{tikz}
\usepackage{booktabs}
\usepackage{multirow}
\usepackage{tabularx}
\usepackage{geometry}
\usepackage{soul}
\usepackage{fancyhdr}
\usepackage{paralist}
\usepackage{setspace}
\usepackage{subcaption}
\usepackage{epstopdf}
\usepackage{authblk}
\usepackage[UKenglish]{babel}
\usepackage[UKenglish]{isodate}
\usepackage[round]{natbib}
\usepackage{csquotes}
\usepackage[title]{appendix}
\usepackage{floatrow}
\usepackage[section]{placeins} 
\usepackage[T1]{fontenc}
\usepackage{floatrow} 
\usepackage[title]{appendix}
\usepackage{caption}
\usepackage{textcomp}
\usepackage{lmodern}
\usepackage{hyphenat}
\usepackage[cal=cm, scr=boondox, frak=euler]{mathalfa} 

\usepackage{ulem}

\usepackage{xr}
\usepackage{titlesec}
\titlelabel{\thetitle \quad}
\usepackage{hyperref}
\hypersetup{
     colorlinks = true,
     allcolors = blue,
 }

\usepackage[]{algorithm2e}
\newcounter{algorithmcustom}

\newcolumntype{Y}{>{\arraybackslash}X}
\newcolumntype{Z}{>{\centering \arraybackslash}X}

\definecolor{orange2}{RGB}{255,185,0}

\usepackage{manyfoot}
\SetFootnoteHook{\hspace*{-1.8em}}
\DeclareNewFootnote{bl}[gobble]
\newcommand{\tickYes}{\checkmark}

\usepackage{amsthm}
\newtheoremstyle{bfremark}%
{}{}%
{}{}%
{\bfseries}{.}%
{ }%
{\thmname{#1}\thmnumber{ #2}\thmnote{ \normalfont (#3)}}
\theoremstyle{bfremark}
\newtheorem{assumption}{Assumption}

\title{Monitoring the Economy in Real Time:\\ Trends and Gaps in Real Activity and Prices}
\date{\today\\[15pt] \color{red}}

\author[1]{Thomas Hasenzagl}
\author[2]{Filippo Pellegrino}
\author[3]{Lucrezia Reichlin}
\author[4]{Giovanni Ricco}
\affil[1]{\it\small University of Minnesota and Federal Reserve Bank of Minneapolis}
\affil[2]{\it\small Imperial College London}
\affil[3]{\it\small London Business School, Now-Casting Economics,  and CEPR}
\affil[4]{\it\small \'Ecole Polytechnique CREST, University of Warwick, OFCE-SciencesPo,  and CEPR}

\begin{document}
\linespread{1.25}

\maketitle
\footnotebl{\scriptsize We are grateful to the 13th International Conference on Computational and Financial Econometrics, European Seminar on Bayesian Econometrics 2021 and Bank of Finland, University of Strathclyde, and Universit\'e de Montr\'eal seminars participants for helpful comments and suggestions. The opinions in this paper are those of the authors and do not necessarily reflect the views of the Federal Reserve Bank of Minneapolis.}
\thispagestyle{empty}

\begin{abstract}

We propose two specifications of a real-time mixed-frequency semi\hyp{}structural time series model for evaluating the output potential, output gap, Phillips curve, and Okun's law for the US. The baseline model uses minimal theory-based multivariate identification restrictions to inform trend-cycle decomposition, while the alternative model adds the CBO's output gap measure as an observed variable. The latter model results in a smoother output potential and lower cyclical correlation between inflation and real variables but performs worse in forecasting beyond the short term. This methodology allows for the assessment and real-time monitoring of official trend and gap estimates.
\end{abstract}
\vspace{0.5cm}

\noindent \small{\textbf{Keywords:} real-time forecasting, output gap, Phillips curve, semi-structural models, Bayesian estimation.}\\ 
\noindent  \small{\textbf{JEL Classification:}  C11, C32, C53, E31, E32, E52.} 

\clearpage
\normalsize
\section*{Introduction}

Economic policy-making requires real-time assessments of the state of the business cycle. Developing such assessments comes with two problems. First, information on important macroeconomic variables is incomplete since statistical agencies publish these data with a publication delay. Second, estimates of long-term trends of these variables are usually imprecise. Still, such estimates are needed to assess the degree of slack in the economy and identify business cycle dynamics (e.g., see discussions in  \citealp{RePEc:tpr:restat:v:84:y:2002:i:4:p:569-583}, \citealp{Watson2007}, and \citealp{RePEc:nbr:nberwo:23580}). 

A large literature on nowcasting has proposed reduced form methods to address the first problem but has focused on tracking growth rates of observed variables, thereby eliminating all long-term trends from the data (see \citealp{RePEc:eee:moneco:v:55:y:2008:i:4:p:665-676} for a seminal contribution)\footnote{Important exceptions are  \cite{RePEc:eee:quaeco:v:54:y:2014:i:2:p:180-193},  \cite{RePEc:eee:quaeco:v:54:y:2014:i:2:p:180-193} and \cite{BarigozziLucianiGap} that develop dynamic factor models for the estimation of the output gap in real-time and  \cite{RePEc:tpr:restat:v:90:y:2008:i:4:p:792-804} who adopt a cointegrated VAR framework to model real-time measures of the output gap and observed time series.}. However, in many situations, the objects of interest are cycles, that is, stationary deviations from low-frequency trends or sometimes even the trends themselves. In those cases, trends must be estimated rather than eliminated by transforming the data to growth rates. Since there are only a limited number of effective observations available in finite samples, estimates of such low-frequency trends are highly uncertain (see \citealp{10.2307/40056492}) and dependent on implicit or explicit assumptions made by the modeller (see \citealp{RePEc:eee:econom:v:95:y:2000:i:2:p:443-462}). Explicit assumptions often come in the form of identifying restrictions on some of the model parameters. Estimating trends and cycles after imposing such restrictions can help reduce uncertainty and improve the predictive performance of the model.\footnote{See, for example, the forecast evaluation in \cite{PriorsLR} that use long-run priors in VAR models, and \cite{hasenzagl2018model} which employs a trend-cycle model.} 

Explicitly incorporating restrictions to identify the unobserved trend and cycle components is sometimes known as {\it structural time series analysis}. It can be thought of as a method to impose conditions on the structure of an unrestricted model such as a VAR. Such restrictions can be derived from economic theory, thus making the approach appealing when the focus of the analysis is on unobserved quantities, such as the output gap, the natural rate of unemployment, or the natural interest rate. Recent applications using small or medium-scale multivariate models are in \cite{RePEc:bin:bpeajo:v:48:y:2017:i:2017-01:p:235-316} (US natural interest rate), \cite{RePEc:ecb:ecbwps:20161966} (Euro Area output gap), and \cite{hasenzagl2018model} (several trends and gaps for the US). The recent work of \cite{NBERw30727} studies the Phillips curve in the US employing this class of models, and offers a methodological discussion.
 
The contribution of our paper is to combine the insights of the nowcasting literature into addressing challenges in real-time forecasting -- how to handle mixed frequency data and unbalanced samples reflecting publication delays and asynchronous data releases --, with those of structural time series analysis. We aim to forecast and track trends and cycles in the economy in real-time. We use a multivariate framework based on the model proposed in \cite{hasenzagl2018model} that includes real and nominal variables and survey data on expectations. Such a framework is coherent with the commonly accepted description of the economy in terms of trends -- such as potential output, the natural rate of unemployment or NAIRU, and trend inflation --, and cycles -- in particular, the output gap and its link to prices via a Phillips curve, and to labor market variables via Okun's law. Our method is also valuable for policy analysis since it can produce nowcasts of observable macroeconomic aggregates and the estimated unobservable components in real-time.

We perform two empirical exercises: First, we estimate two versions of our model with US data from January 1985 until December 2019 and compare the estimated trends and cycles. The two versions are an {\it undisciplined} model that we directly estimate from macroeconomic aggregates, and a {\it tracking} model that is identical to the undisciplined model except that it also incorporate the Congressional Budget Office's (CBO) measure of the output gap as an observable variable. We adopt the CBO output gap measure since it is the standard reference for academic and policy work in the United States. It results from years of experience combining a production-function approach with expert judgment, and is an important benchmark for output gap estimates. 

In both the undisciplined and the tracking models, the estimates of the structural components are informed by the data and multivariate restrictions that connect real, labour market, and nominal variables via Okun's law and the Phillips curve. In the undisciplined model, the estimate of the trend's relative variability and the GDP cycle is determined solely by the data, our identifying assumptions, and Bayesian priors, which we set as close to being uninformative as possible. Conversely, in the tracking model, beliefs about the output gap enter via the CBO's cyclical measure, which has implications for the multivariate links between output, prices, and expectations. The two models provide equally plausible alternative ways to fit the data. Comparing results across them can shed light on the implications of the CBO view on the output gap and help assess its validity.

Our results point to many similarities between the two models but also significant differences. The tracking model reads the US economy as having been constantly below potential since the 2001 recession, reflecting the implicit view on output potential being unaffected by events since then. Conversely, the undisciplined model estimates an almost symmetrical output gap fluctuating around a trend whose slope has declined since 2001. The divergent views are reflected in differences in the estimate of the slope of the Phillips curve, especially since 2001. The undisciplined model identifies a larger common cycle between inflation and real variables and implies, on average, a slightly larger cyclical component in inflation related to the output gap.

In our second empirical exercise, we compare the forecasting performance of the two models using real-time data vintages that mimic the actual release and revision patterns of the macroeconomic variables in our model. The forecast evaluation shows that the undisciplined model outperforms the tracking model at the business cycle frequency, but does not offer an advantage in short-term forecasting. This is true for GDP, labor market variables, and inflation. This result is surprising since it implies that including additional information in the form of the CBO's gap does not improve the model's forecasting accuracy. It points to the fact that the statistical signal extraction performed by the model uncovers significant correlations between inflation and real variables at business cycle frequency, which improves the out-of-the-sample forecasting performance. This result also suggests that the CBO's implicit view of a stable potential output may lead to a misinterpretation of the cyclical components and hence the medium-term inflationary pressures.

As a by-product of the analysis, the mixed frequency feature of our model allows us to estimate both a monthly version of the CBO output gap and a sequence of nowcasts for it. A real-time evaluation of the CBO output gap nowcast during the COVID sample shows that the revisions of the output gap in the undisciplined model are smaller than those in the implied monthly CBO output gap. 

The paper is organised as follows. We discuss the methodology in Section \ref{sec:methodology} and introduce the {\it undisciplined} and {\it tracking} models. Section \ref{sec:results}  discusses in-sample and out-of-sample results, while Section \ref{sec:real-time}  provides a real-time analysis of the output gap during the Covid period. The last section concludes. The Online Appendix provides details on the Bayesian estimation of the model and additional results for all of the models discussed in the paper.

\section{Trends and cycles in the US economy}\label{sec:methodology}

\subsection{A stylised representation of the economy}

Our modelling approach is motivated by a conventionally accepted stylised representation of economic variables in terms of trends and cycles. Output is generally described as fluctuating around a long-run trend (potential output), that is driven by demographic trends, capital accumulation and technological innovation. The trend component is often modelled as a non-stationary unit root process. Different shocks can push output above or below its potential. The fluctuations off equilibrium are defined in term of an output gap, often modelled as an AR(p) stationary process, that can be seen as the `primitive' measure of business cycles. Such a stylised description can be formulated as
\begin{eqnarray}
y_{t} &=& \tau_{t}^{y} + \hat y_t^{gap} = \tau_{t}^{y} + \psi_{t}^{gap} \ , \\
\psi_{t}^{gap} &=& \rho(L)\psi_{t-1}^{gap} + v_t \ , \\
\tau_{t}^{y} &=& \mu + \tau_{t-1}^{y} + \eta_t \ ,
\end{eqnarray}
where $\tau_{t}$ and $\psi_{t}^{gap}$ are the output potential and the output gap. The first is a unit root process with a drift $\mu$, and the second an autoregressive stationary process. $v_t$ and $\eta_t$ are i.i.d. innovations to the two components, while $L$ is the lag operator.

Slack in the economy is reflected into labour market variables via the Okun's law, with unemployment oscillating around a long-run equilibrium level ($\tau_{t}^{u}$). This is the unemployment rate consistent with output at its potential and no inflationary pressure, and is commonly referred to as the non-accelerating inflation rate of unemployment (NAIRU):
\begin{equation}
u_{t} = \tau_{t}^{u} + \hat u_t^{gap} = \tau_{t}^{u} + \gamma_u \hat y_t^{gap} \ .
\end{equation}

Prices fluctuate at business cycles frequencies around an underlying trend inflation, $\tau_{t}^{\pi}$, that is anchored by the inflation target of a credible central bank, and is reflected into the long-run expectations of agents. Deviation from trend inflation are either due to the transmission of cyclical pressure to prices (the Phillips curve), or to short-lived idiosyncratic disturbances, $\psi_{t}^{epc}$, possibly related to energy prices that directly enter the basket of consumption, i.e.
\begin{equation}
\pi_{t} = \lim_{h \to \infty} E_t \pi_{t+h} +\hat \pi_t^{gap} + \psi_{t}^{epc}  = \tau_{t}^{\pi} + \gamma_\pi \hat y_t^{gap} + \psi_{t}^{epc}  \ .
\end{equation}

Such a description of the economy can be summarised by a model of idiosyncratic and common components, capturing the long-run behaviour of the variables and their business-cycle fluctuations,
\begin{equation}
\begin{pmatrix}
y_{t} \\ 
u_{t} \\ 
\pi_{t} \\ 
\end{pmatrix} =
\begin{pmatrix}
1 & 0 \\ 
\gamma_u  & 0 \\ 
\gamma_\pi  & 1
\end{pmatrix}
\begin{pmatrix}
\psi_{t}^{gap} \\ 
\psi_{t}^{epc} 
\end{pmatrix}
+ 
\begin{pmatrix} 
\tau_{t}^{y}  \\ 
\tau_{t}^{u} \\ 
\tau_{t}^{\,\pi} \\ 
\end{pmatrix}
.
\label{model_nut} 
\end{equation}
While such a description is in line with textbooks and, possibly, the policymakers' view, it is too stylised for empirical analysis. To bring the model to the data we need to introduce lags to account for heterogeneous dynamics across variables, and idiosyncratic shocks reflecting measurement errors or wedges with the theory. Indeed, it is well know that unemployment and prices reacts with lags to the slack in the economy and that potentially several idiosyncratic components can distort both formation of economic expectations and the dynamics of the variables themselves. In the next sections, we describe the empirical version of the model and the econometric methodology.

\subsection{A mixed-frequency trend-cycle framework}

Our empirical framework adopts and generalises the simple model described in the previous section to capture the joint dynamics of real activity -- i.e. output, employment and unemployment rate --, nominal variables -- i.e. consumer price inflation and oil prices --, and expectations -- i.e. professional forecasts of inflation and output, consumers' expectations of inflation. 

The multivariate relationships are expressed  in terms of common cycles and trends that are meant to capture structural components and their dynamics, plus a number of variable-specific components that absorb idiosyncratic shocks and measurement errors. The model incorporates variables at the frequency at which they are published by the statistical offices. \autoref{tab:data} summarises the variables, the frequencies at which they enter the model, and the common trends and cyclical components.

\begin{table}[t!]
	\footnotesize
	\centering
	\begin{tabularx}{\textwidth}{@{}lcccccc@{}}
		\toprule
		\multirow{2}{*}{\textbf{Variable name}}        	&  \multirow{2}{*}{\textbf{Label at time t}}  &  \multirow{2}{*}{\textbf{Frequency}} & \multicolumn{4}{c}{\textbf{Loads on}} \\
		&                    			 	&& {\it BC}          & {\it EPC}             & {\it GDP trend}  & {\it Trend} $\pi$\\
		\midrule
		CBO: cycle of real GDP$^{*}$          & $gap_{t}^{\,cbo}$      & Q          & \tickYes           &            		&   & \\
		Real GDP                     	 & $y_{t}$      & Q          & \tickYes           &            			&  \tickYes  & \\
		SPF: expected real GDP   & $F_t^{y} \pi_{t+12}$      & Q          & \tickYes           &            			&  \tickYes  & \\
		Unemployment rate        & $u_{t}$      & M          & \tickYes           &            			&        		  & \\
		Employment                   & $e_{t}$      & M          & \tickYes           &            			&        		  & \\
		WTI spot oil price           & $oil_{t}$      & M         &            				& \tickYes  		&        		  & \\
		CPI                         		 & $\pi_{t}$  & M       & \tickYes           & \tickYes  		 &        		   & \tickYes \\                    
		SPF: expected inflation    & $F_t^{spf} \pi_{t+12}$     & Q         & \tickYes          & \tickYes 	 		&        		  & \tickYes \\
		UoM: expected inflation  & $F_t^{uom} \pi_{t+12}$     & M        & \tickYes          & \tickYes  		  &        			& \tickYes \\
		\bottomrule
	\end{tabularx}
	\caption{US data and common components}
	\floatfoot{\textbf{Notes:} Data used in the trend-cycle model. All data is in levels, except for CPI which is in YoY (\%). `UoM: expected inflation' is the University of Michigan, 12-months ahead expected inflation. `SPF: expected inflation' is the Survey of Professional Forecasters, 4-quarters ahead expected inflation rate. Data includes observations from Jan-1985 to Dec-2019. (*) Used in the tracking model only.}
	\label{tab:data}
\end{table}

A number of assumptions underpins our framework. Let us start from the trends.

\begin{assumption}[\bf Output potential] The output potential is the stochastic trend driving output in the long run, and is the common trend between real GDP and expected real GDP. In the spirit of \cite{beveridge1981new}, it coincides with the long-run forecast of output implied by the model.
\end{assumption}

\begin{assumption}[\bf Labor market trends]
Employment and the unemployment rate have each their own trend defined as their long-run forecast. We denote them as as $\tau_{t}^{e}$ and $\tau_{t}^{u}$, respectively. $\tau_{t}^{u}$ is the estimate of the non-accelerating inflation rate of unemployment (NAIRU).
\end{assumption}

\begin{assumption}[\bf Trend inflation]
Trend inflation, $\tau_t^\pi$, is the common trend shared by inflation itself and inflation expectations. It is also the long-run model-based forecast of inflation.
\end{assumption}

The cyclical components are modelled as stationary stochastic cycles, under the following set of assumptions.

\begin{assumption}[\bf Output gap] 
In the spirit of \cite{RePEc:nbr:nberbk:burn46-1}, the output gap $\psi_{t}^{gap}$ defined as an economy-wide stationary stochastic  component common to all real variables, labor market variables, inflation, and survey expectations. It informs the price gap via the Phillips curve, and the unemployment gap via the Okun's law. Both relationships are modelled as moving averages of output gap realisations over the previous three months.\footnote{It is worth observing that the all variables, except for real GDP and the CBO's output cycle, are connected to the output gap with a lag polynomial. This is to allow the model to nest, under parametric restrictions, the case of rational expectations, as discussed in \cite{hasenzagl2018model}.}
\end{assumption}

We also consider a second common stationary component, which we call the ``energy price component'' that captures the direct effect of energy shocks into headline inflation.

\begin{assumption}[\bf Energy price component] The energy price component $\psi_{t}^{epc}$ is a stationary stochastic common cyclical component connecting oil prices, inflation, and inflation expectations.
\end{assumption}

A number of idiosyncratic stationary components absorbs various forms of noise which could distort the empirical estimates of the structural relationships.

\begin{assumption}[\bf Idiosyncratic stationary components]
All variables have an idiosyncratic stationary component, $\psi_{i,t}$, which absorbs different sources of idiosyncratic dynamics such as idiosyncratic shocks, non-classic measurement error, differences in definitions, and other sources of noise.
\end{assumption}

Finally, we introduce a number of non-stationary components to capture persistent time-varying biases in survey data.

\begin{assumption}[\bf Bias in Expectations] Agents' expectations can deviate from a rational forecast due to time-varying bias -- respectively $\mu_{t}^{spf, y}$, $\mu_{t}^{spf, \pi}$ for the professional forecasters' and  $\mu_{t}^{uom, \pi}$ for consumers' expectations. The bias terms are modelled as stochastic random walk components.
\end{assumption}

While, in the general form of the model, we can allow for these biases to affect both consumers' and professionals' expectations, empirically only consumers' expectations exhibit persistent biases (see \citealp{CoibionGoro2015}, for a discussion). In line with this observation, in the empirical section of this paper, we set $\mu_{t}^{spf, \pi}$ to zero (the professional forecasts for inflation (SPF) are `on trend' at all times), and we only allow the constant $\mu^{spf, y}$ to account for measurement differences in the expected output trend, possibly due to measurement and aggregation issues.\\

Taken together these assumptions imply a representation of the variables of interest of the form
\begin{align}
\resizebox{0.93\hsize}{!}{
$\begin{pmatrix} cycle_{t}^{\,cbo} \\ 
y_{t} \\ 
F_t^{spf} y_{t+12} \\ 
u_{t} \\ 
e_{t} \\ 
oil_{t} \\ 
\pi_{t} \\ 
F_t^{spf} \pi_{t+12} \\ 
F_t^{uom} \pi_{t+12} 
\end{pmatrix} =
\underbrace{
\begin{pmatrix} 
\sum_{j=0}^2 L^j & 0 \\ 
\sum_{j=0}^2 L^j  & 0 \\ 
\sum_{j=0}^3 \gamma_{3,j} L^j & 0\\ 
\sum_{j=0}^3 \gamma_{4,j} L^j & 0\\ 
\sum_{j=0}^3 \gamma_{5,j} L^j & 0\\ 
0 & 1 \\ 
\sum_{j=0}^3 \gamma_{7,j} L^j  & \delta_{7} \\ 
\sum_{j=0}^3 \gamma_{8,j} L^j  & \delta_{8} \\ 
\sum_{j=0}^3 \gamma_{9,j} L^j  & \delta_{9} 
\end{pmatrix}
\begin{pmatrix}
\psi_{t}^{gap} \\ 
\psi_{t}^{epc}
\end{pmatrix}
+
\begin{pmatrix}
\sum_{j=0}^2 L^j \psi_{1,t} \\ 
\sum_{j=0}^2 L^j \psi_{1,t} \\ 
\psi_{3,t} \\ 
\psi_{4,t} \\ 
\psi_{5,t} \\ 
\psi_{6,t} \\ 
\psi_{7,t} \\ 
\psi_{8,t} \\ 
\psi_{9,t}
\end{pmatrix}
}_\text{Common \& Idiosyncratic Cycles}
+ 
\underbrace{
\begin{pmatrix}
0 & 0 & 0 & 0 & 0 & 0  & 0\\ 
\sum_{j=0}^2 L^j & 0 & 0 & 0 & 0 & 0  & 0\\ 
3 & 1 & 0 & 0 & 0 & 0 & 0\\ 
0 & 0 & 1 & 0 & 0 & 0 & 0\\ 
0 & 0 & 0 & 1 & 0 & 0 & 0\\ 
0 & 0 & 0 & 0 & 1 & 0 & 0\\ 
0 & 0 & 0 & 0 & 0 & 1 & 0\\ 
0 & 0 & 0 & 0 & 0 & 1 & 0\\ 
0 & 0& 0 & 0 & 0 & 1  & 1
\end{pmatrix}
\begin{pmatrix} 
{\color{red}\tau^{y}}  \\ 
\mu_{t}^{spf, y} \\ 
\tau_{t}^{u} \\ 
\tau_{t}^{e} \\ 
\tau_{t}^{oil} \\ 
\tau_{t}^{\,\pi} \\ 
\mu_{t}^{uom, \pi} 
\end{pmatrix}
}_\text{Trends \& Biases}
,$}
\label{model}
\end{align}
where $L$ is the lag operator.\\

We consider two specifications: 
\begin{enumerate}
\item the tracking model that incorporates the CBO measure of the cycle of GDP as an observed quarterly measure of the output gap (as in \autoref{model}); 
\item the undisciplined model that is as specified in \autoref{model} but does not include $cycle_{t}^{\,cbo}$. Hence the output gap is an additional unobserved component that the model has to estimate. 
\end{enumerate}

To complete the state-space representation of the model, we specify the dynamic equations governing the evolution of the unobserved components over time following the approach of \cite{harvey1985trends}.
	
\begin{assumption}[\bf State dynamics]
The stationary cycles are all modelled as ARMA(2,1) stochastic processes, with coefficients restricted to produce stationary oscillations of defined periodicity. The trends are random walks. Specifically, potential output and trend employment are random walks with drift, and all the remaining trends are driftless random walks. All of the processes have mutually orthogonal stochastic innovations.
\end{assumption}

It is worth noticing that ARMA(2,1) processes displays pseudo-cyclical behaviour and can be conveniently written in a VAR(2) representation as
\begin{align}
    &\widehat \psi_{t} = \rho \cos(\lambda) \widehat \psi_{t-1} + \rho \sin(\lambda) \widehat \psi^*_{t-1} + v_{t} \ , \label{output_gap1} \\
    &\widehat \psi^*_{t} = -\rho \sin(\lambda) \widehat \psi_{t-1} + \rho \cos(\lambda) \widehat \psi^*_{t-1} + v^*_{t}  \nonumber \ ,
\end{align}  
where the parameters $0 \leq \lambda \leq  \pi$ and $0 \leq \rho \leq 1$ can be interpreted, respectively, as the frequency and the damping factor on the amplitude of the cycle (the process is stationary for $\rho < 1$). $\widehat \psi^*_{t}$ is an auxiliary cycle that support the VAR(2) representation, and $v_{t}$ and $v^*_{t}$ are uncorrelated white noise disturbances \citep[see][]{harvey1990forecasting}.\footnote{It is straightforward to show that the model can be rewritten as
$$
(1-2\rho \cos(\lambda)L + \rho^2 L^2) \widehat \psi_{t} = (1 - \rho \cos(\lambda)L)v_{t} + (\rho \sin(\lambda)L)v^*_{t} \ .
$$
Hence, under the restriction $\sigma^2_{v} = 0$, the solution of the model is an AR(2), otherwise an ARMA(2,1). The intuition for the use of the auxiliary cycle is closely related to the standard multivariate AR(1) representation of univariate AR(p) processes.} The disturbances make the cycle stochastic rather than deterministic. 

\subsection{Mixed-frequency aggregation}

The model incorporates data at different frequencies (see \autoref{tab:data}). Specifically, quarterly indicators are treated as monthly data with missing observations for which the model has to deliver estimates, employing a set of restrictions similar to those proposed in \cite{mariano2003new}.

\begin{assumption}[\bf Aggregation rules for the CBO cycle and real GDP] \label{assumption:aggregation_GDP}
For the CBO cycle and real GDP, the quarterly data are linked to latent monthly figures (denoted with the use of a tilde) as
\begin{align*}
	&cycle_{t}^{\,cbo} = (1+L+L^2) \, \widetilde{\,cycle}_{t}^{\,cbo}, \\
	&y_{t} = (1+L+L^2) \, \tilde{y}_{t},
\end{align*}
where
\begin{align}
	&\widetilde{\,cycle}_{t}^{\,cbo} = \tilde{y}_{t} - \tau_{t}^{y}, \label{eq:aggregation_1} \\
	&\tilde{y}_{t} = \psi_{t}^{gap} + \psi_{1,t} + \tau_{t}^{y}, \label{eq:aggregation_2}
\end{align}
for any $t$. This aggregation approach is standard and used in several papers including \cite{RePEc:eee:moneco:v:55:y:2008:i:4:p:665-676} and \cite{banbura2014maximum}. 
\end{assumption}

\begin{assumption}[\bf Aggregation rules for expectational data] 
Professionals' expectations for real GDP and inflation are aggregated in different ways. At any $t$ we have\footnote{We use $F_t x_{t+h}$ to indicate survey expectations at time $t$ for a variable $x_{t+h}$ to distinguish them from mathematical expectations, $E_t x_{t+h}$.}
\begin{align*}
	F_t^{spf} y_{t+12} = F_t^{spf} \tilde{y}_{t+12} + F_t^{spf} \tilde{y}_{t+11} + F_t^{spf} \tilde{y}_{t+10},
\end{align*}

\noindent since $F_t^{spf} y_{t+12}$ is quarterly and follows equivalent aggregation rules to the ones in \autoref{assumption:aggregation_GDP}. In other words, we link the professional expectation $F_t^{spf} y_{t+12}$ with implied predictions for monthly real GDP figures computed with the same conditioning set (i.e., the one available at time $t$).\\

In the case of inflation, the professional expectations do not need to be linked with latent figures measuring inflation at a higher frequency than the one at which it is sampled, since the actual data is monthly already. Therefore, we simply consider $F_t^{spf} \pi_{t+12}$ as the end-of-month one-year ahead forecast for inflation.\\

It follows that we need to enforce a mixed-frequency aggregation rule only for the real GDP professional expectations. Not knowing the exact prediction rule followed by professional forecasters, we cannot recover the expectations for the latent monthly output figures in their entirety. However, we know that the persistent component of these expectations should be linked to the trends of real GDP. For simplicity, we assume them to be the same and, thus, enforce an aggregation rule according to which the trend of $F_t^{spf} y_{t+12}$ is considered as $(1+L+L^2) \tau_{t+12}^{y}.$ Since the trend is a random walk with drift, the trend can also be written as the sum of  $3 \tau_{t}^{y}$ plus a time-invariant drift. The time-invariant drift, denoted as $\mu^{spf,y} \equiv \mu_t^{spf,y}$ for every $t$, is estimated and so are the loadings.
\end{assumption}

\subsection{Bayesian estimation}

The model is casted in a in linear state-space form and estimated with Bayesian techniques, employing an Adaptive Metropolis-Within-Gibbs algorithm (details are provided in the Online Appendix, in \autoref{AMWG}). We adopt the simulation smoother of \cite{durbin2002simple} along with the \cite{jarocinski2015note}'s modification to condition our estimates of cycles and trends on the full sample.

Data of each variable are normalised by dividing them for the standard deviation of their first-differences.\footnote{As discussed in \cite{hasenzagl2018model} this normalisation gives set data on a similar scale and provides better mixing in the Metropolis algorithm.} To deal with missing observations, we employ a Kalman filter approach \citep[see, as a reference, the discussion in][]{shumway1982approach}, and reconstruct the data on the basis of the information available at each point in time.

\section{Trends and gaps in the US economy}\label{sec:results}

How do the two models read the business cycle fluctuations and trends in the US economy? We start in this section by assessing in-sample results from January 1985 to December 2019 and using fully revised data. The following section provides a real-time appraisal of the two models' performances.

\begin{figure}[t!]
    \centering
	{\bf \footnotesize Tracking Model} \hspace{0.24\textwidth} {\bf \footnotesize  Undisciplined Model}
	\vfill

    \begin{subfigure}[b]{0.24\textwidth}
        \centering
        \includegraphics[width=\textwidth]{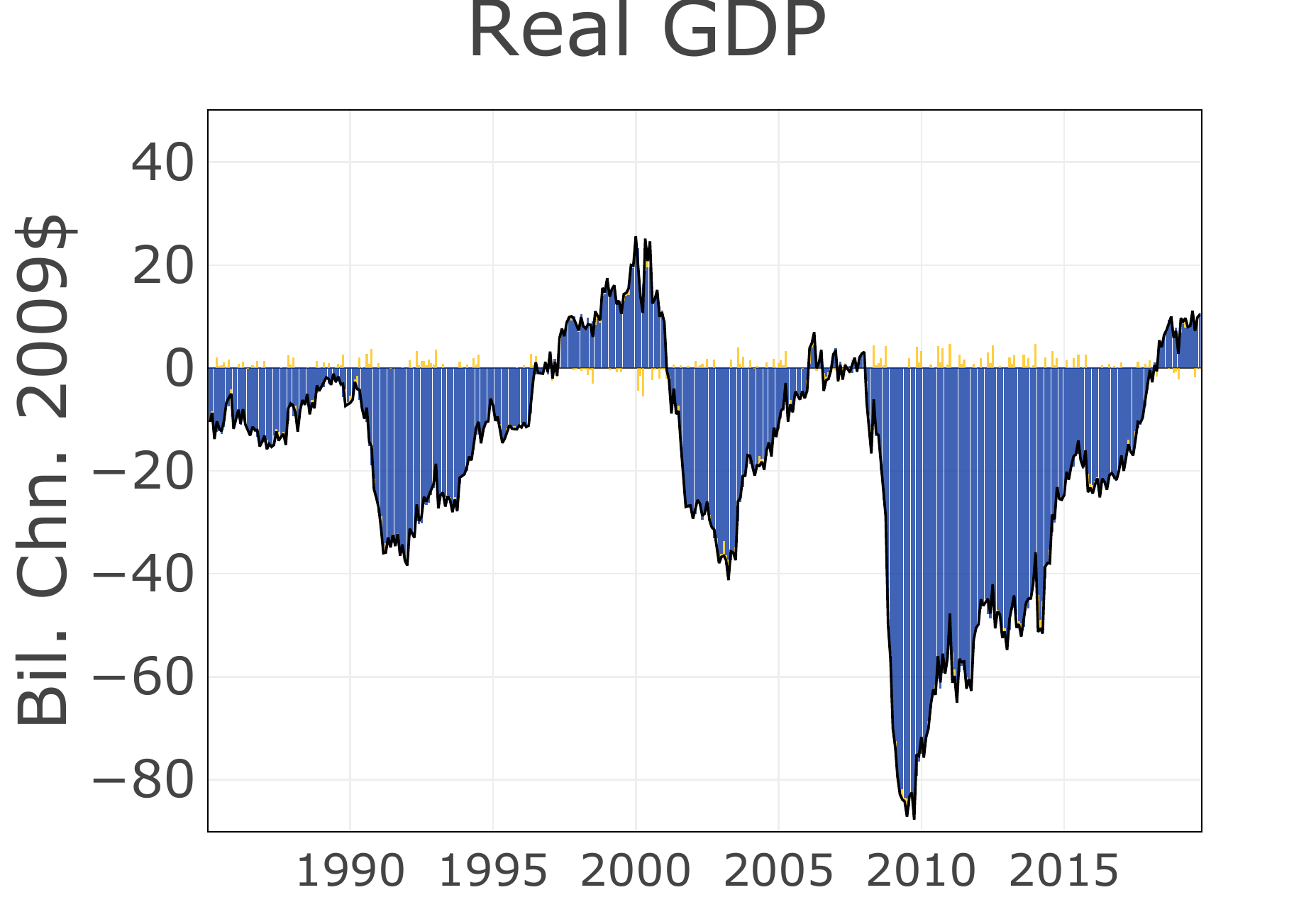}
    \end{subfigure}
    \hfill
    \begin{subfigure}[b]{0.24\textwidth}
        \centering
        \includegraphics[width=\textwidth]{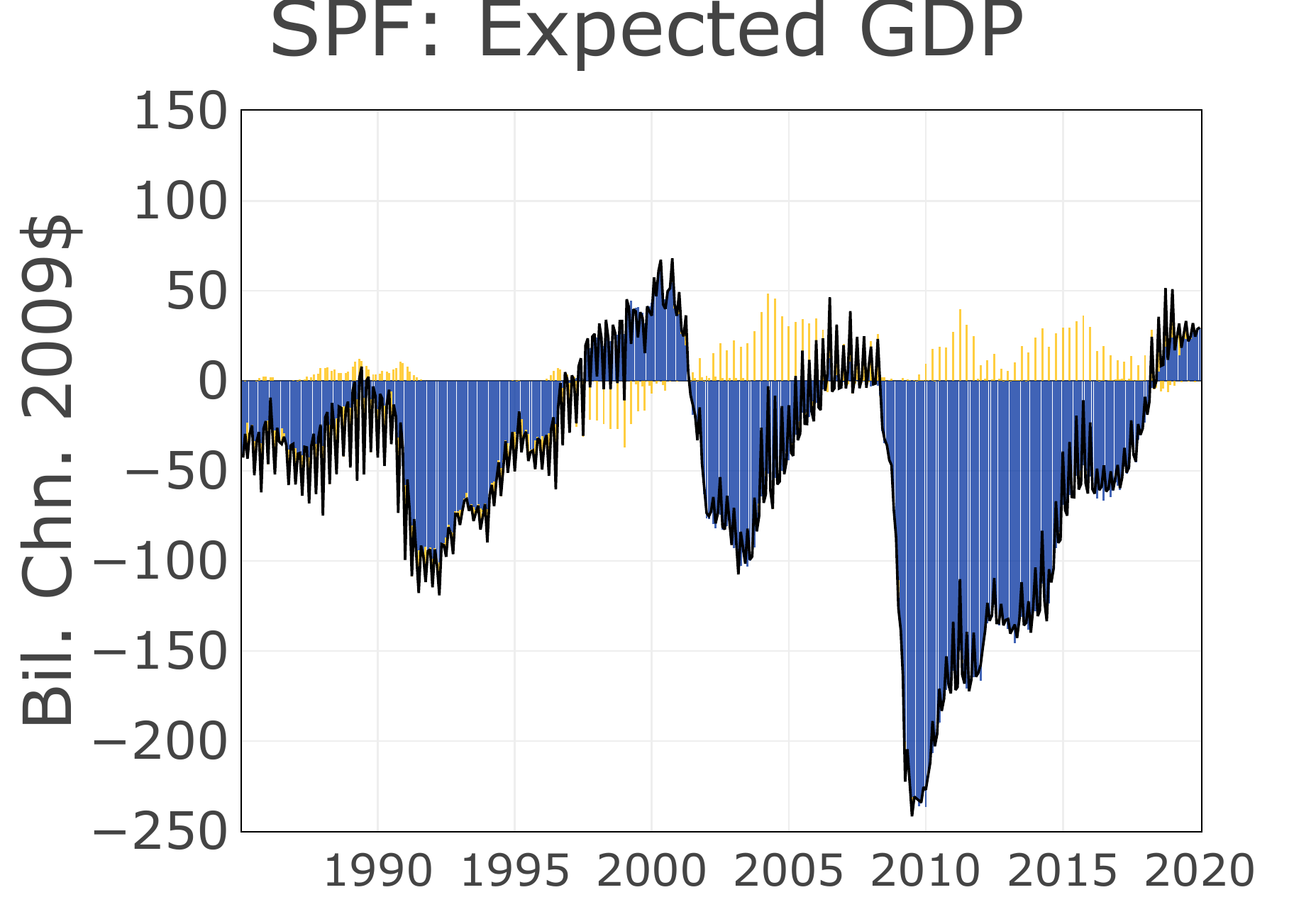}
    \end{subfigure}
    \hfill
    \vrule\
    \begin{subfigure}[b]{0.24\textwidth}
        \centering
        \includegraphics[width=\textwidth]{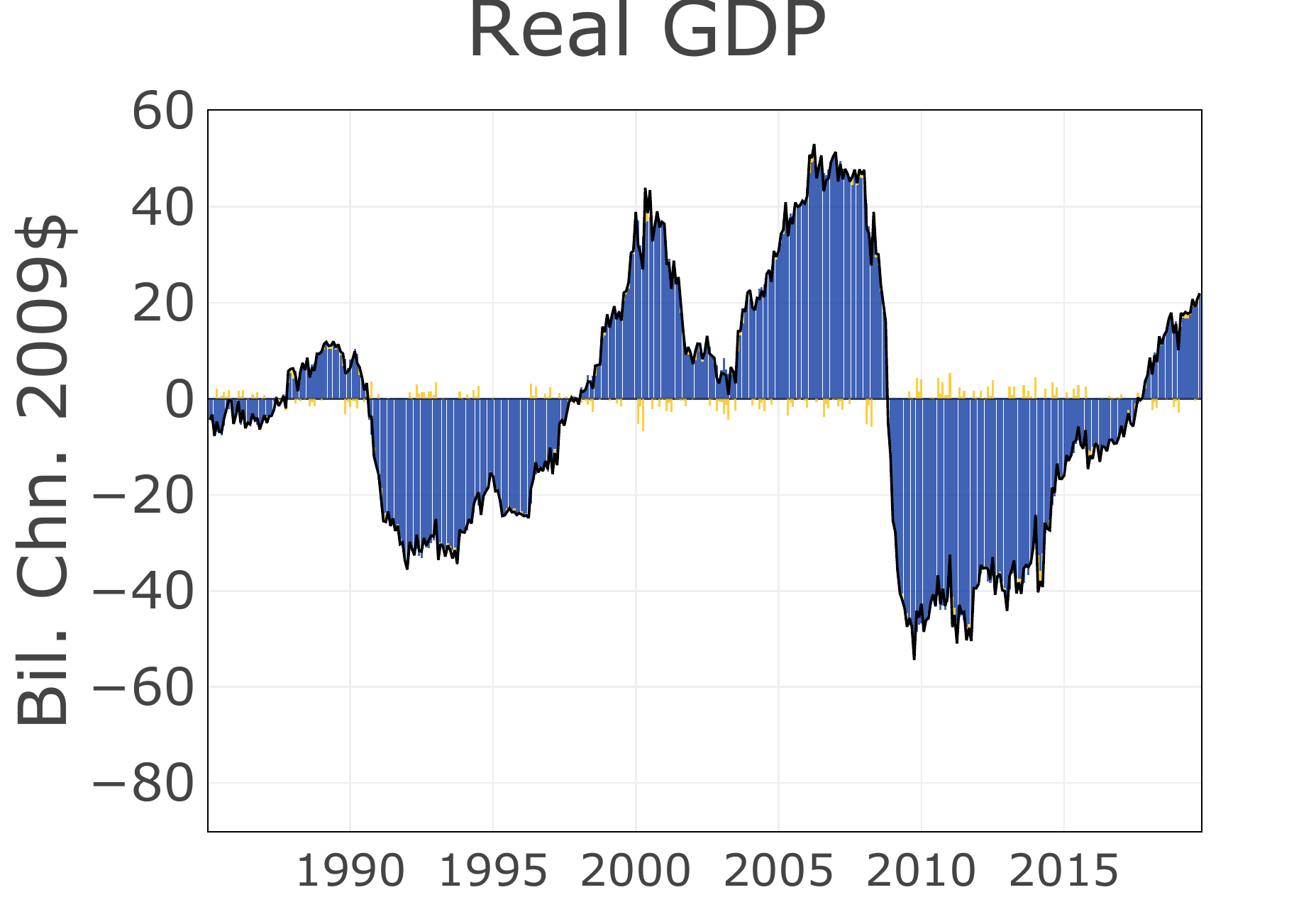}
    \end{subfigure}
	\hfill
    \begin{subfigure}[b]{0.24\textwidth}
        \centering
        \includegraphics[width=\textwidth]{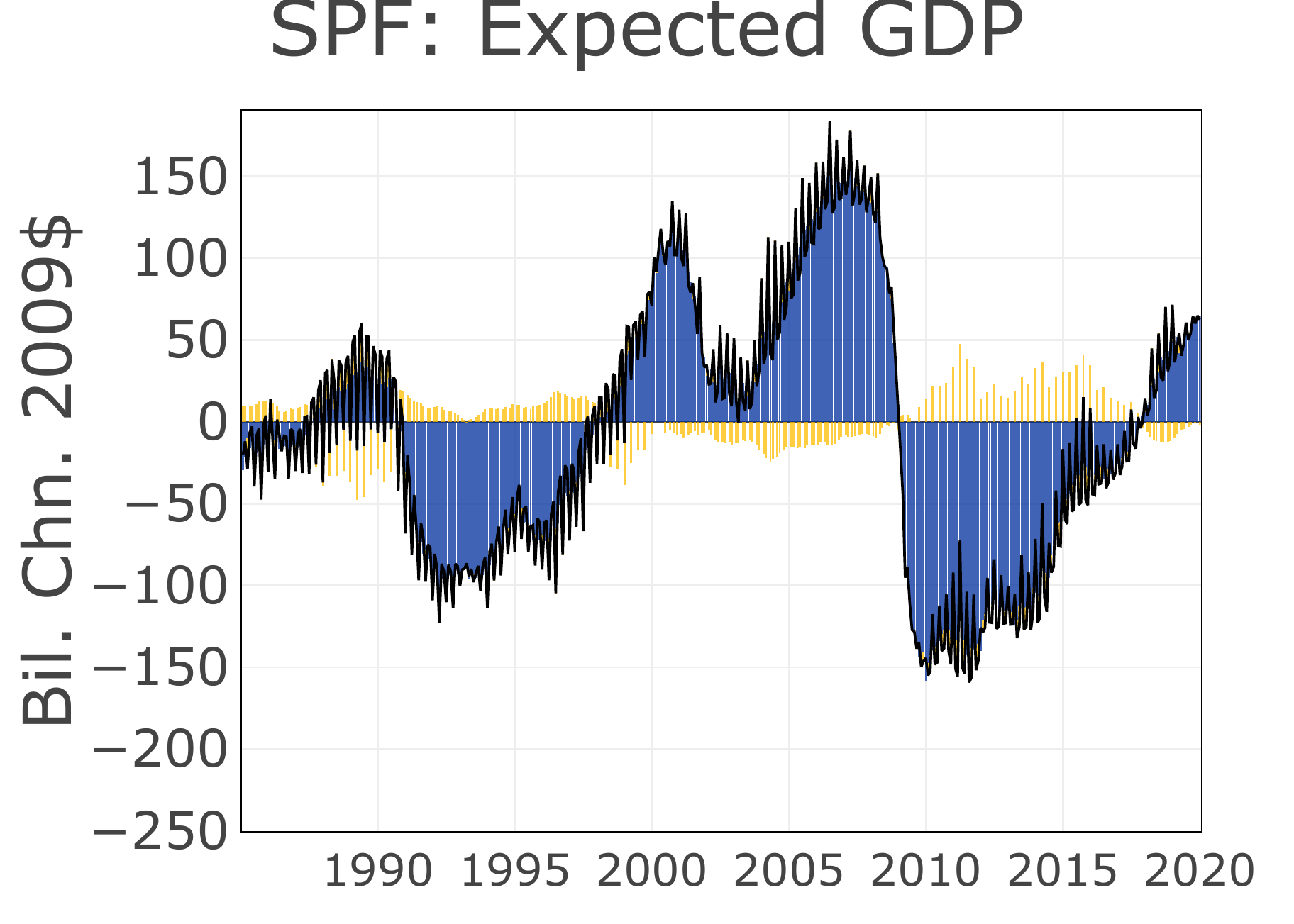}
    \end{subfigure}

    \begin{subfigure}[b]{0.24\textwidth}
        \centering
        \includegraphics[width=\textwidth]{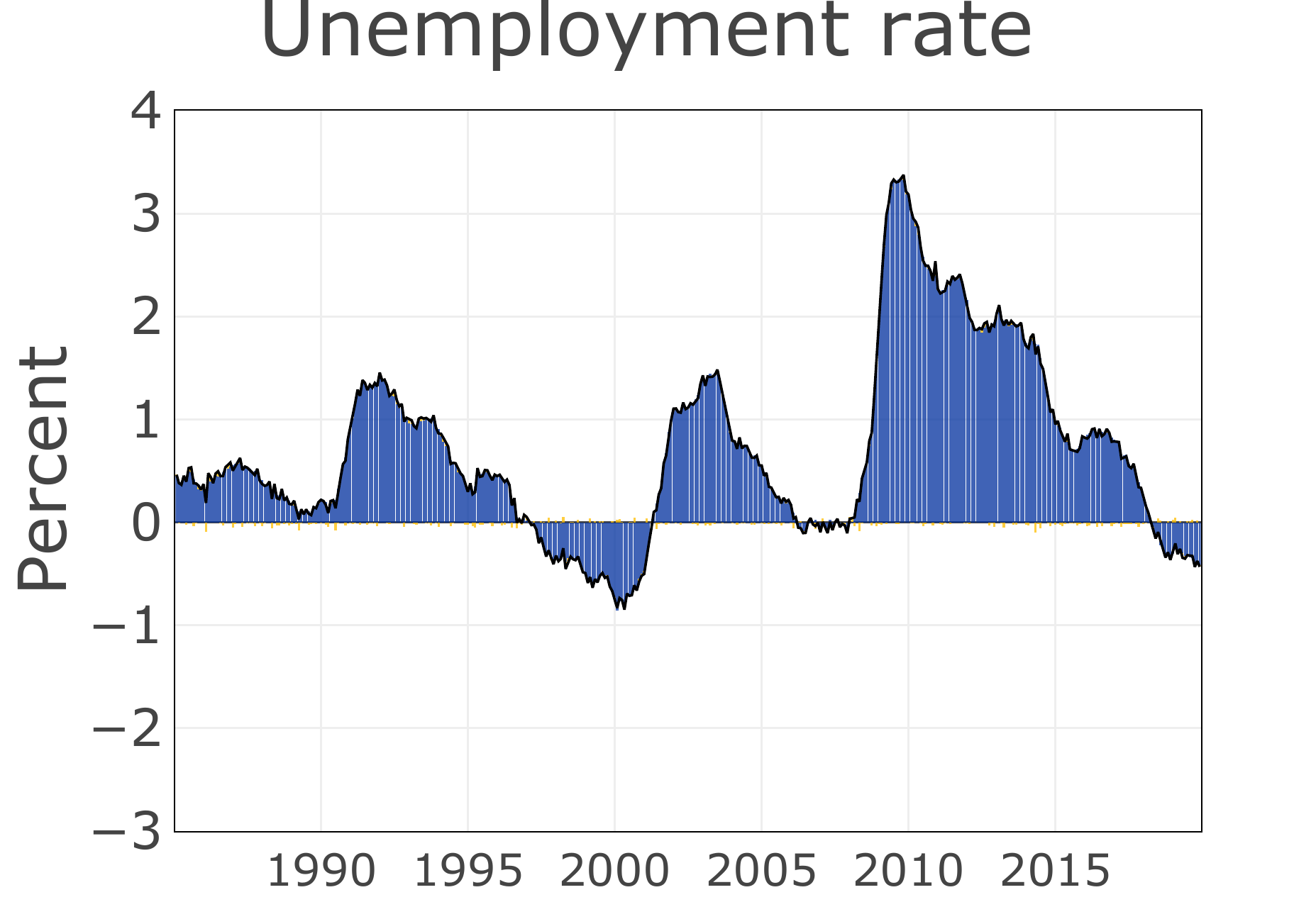}
    \end{subfigure}
    \hfill
    \begin{subfigure}[b]{0.24\textwidth}
        \centering
        \includegraphics[width=\textwidth]{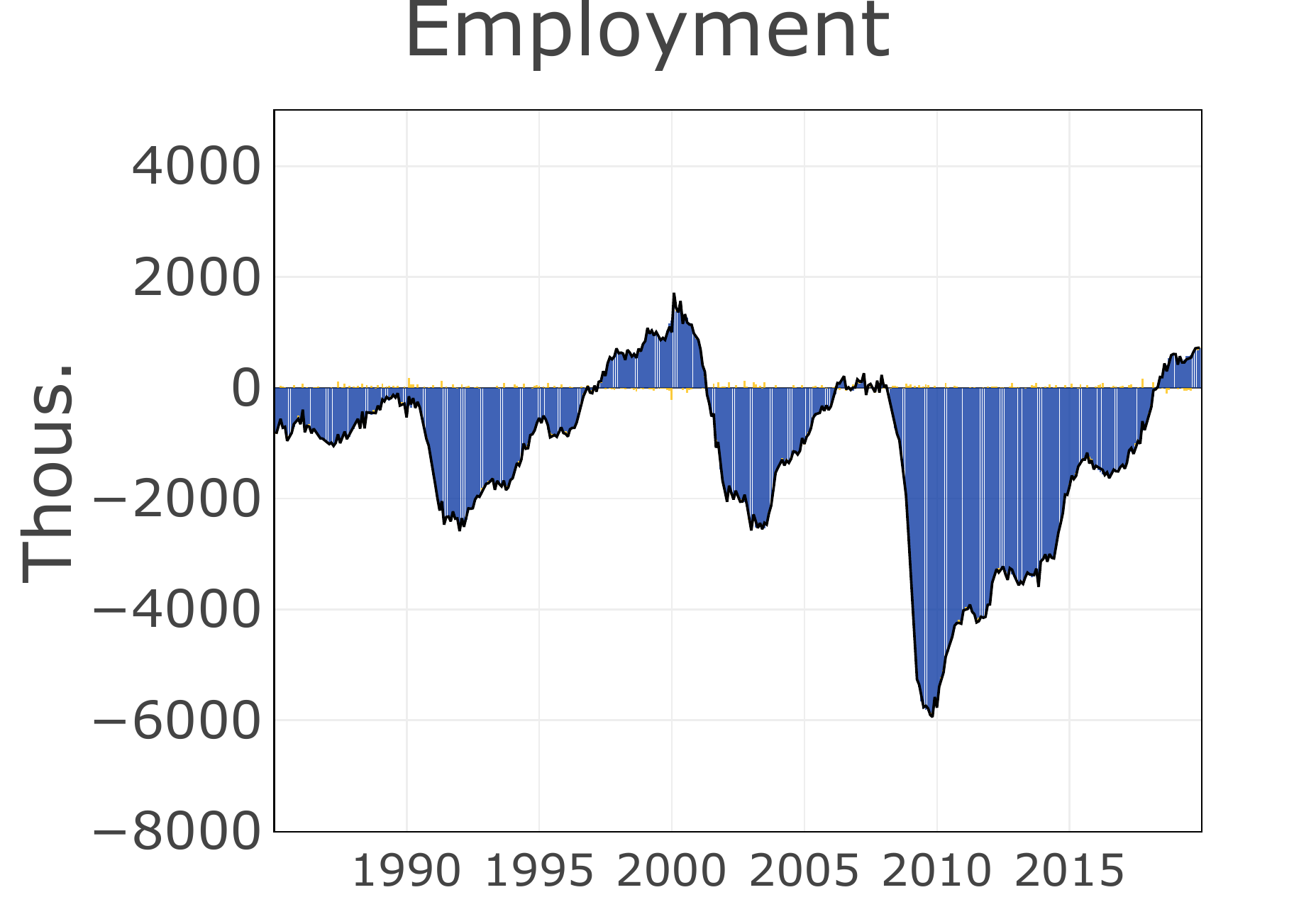}
    \end{subfigure}
    \hfill
    \vrule\
    \begin{subfigure}[b]{0.24\textwidth}
        \centering
        \includegraphics[width=\textwidth]{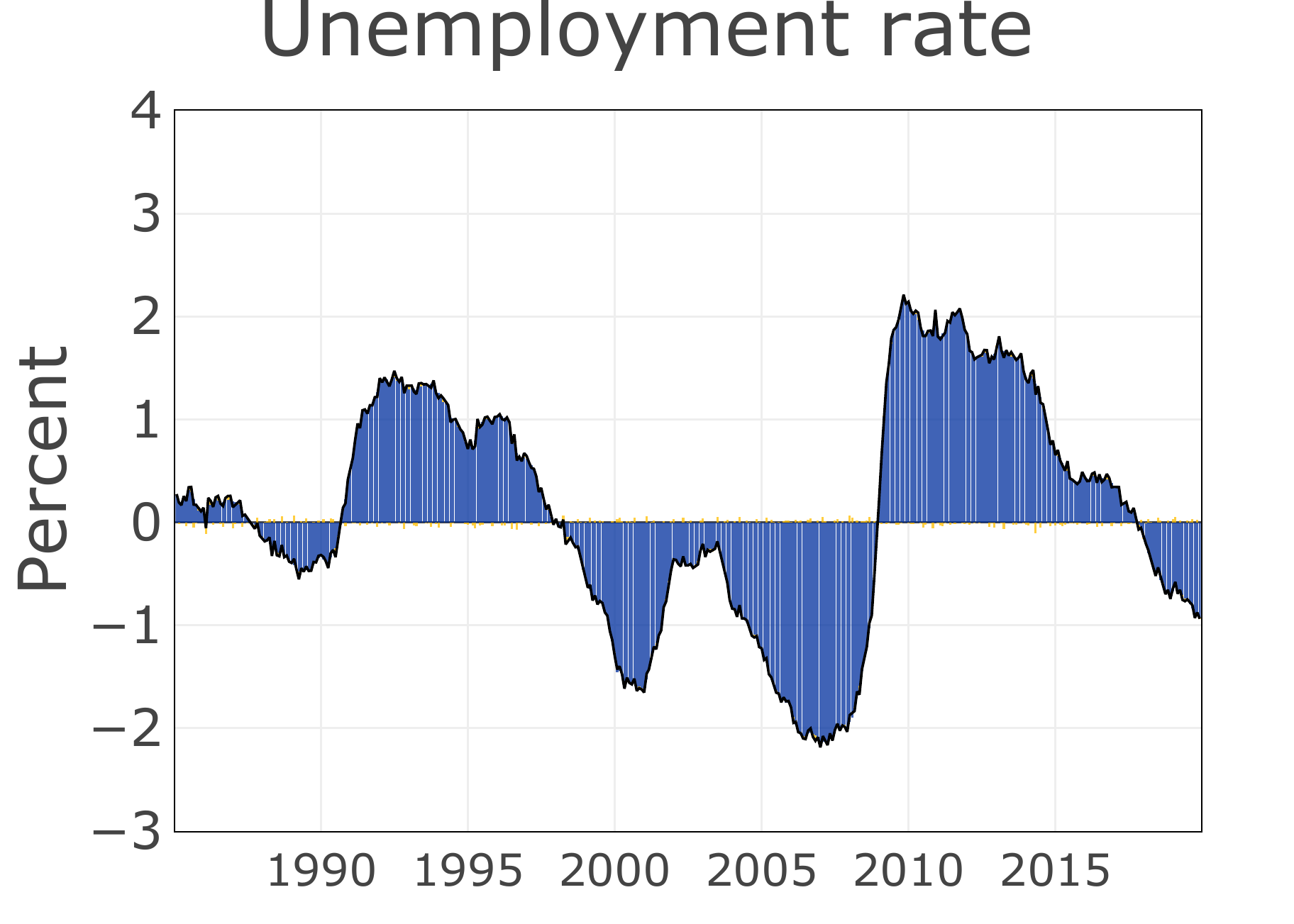}
    \end{subfigure}
	\hfill
    \begin{subfigure}[b]{0.24\textwidth}
        \centering
        \includegraphics[width=\textwidth]{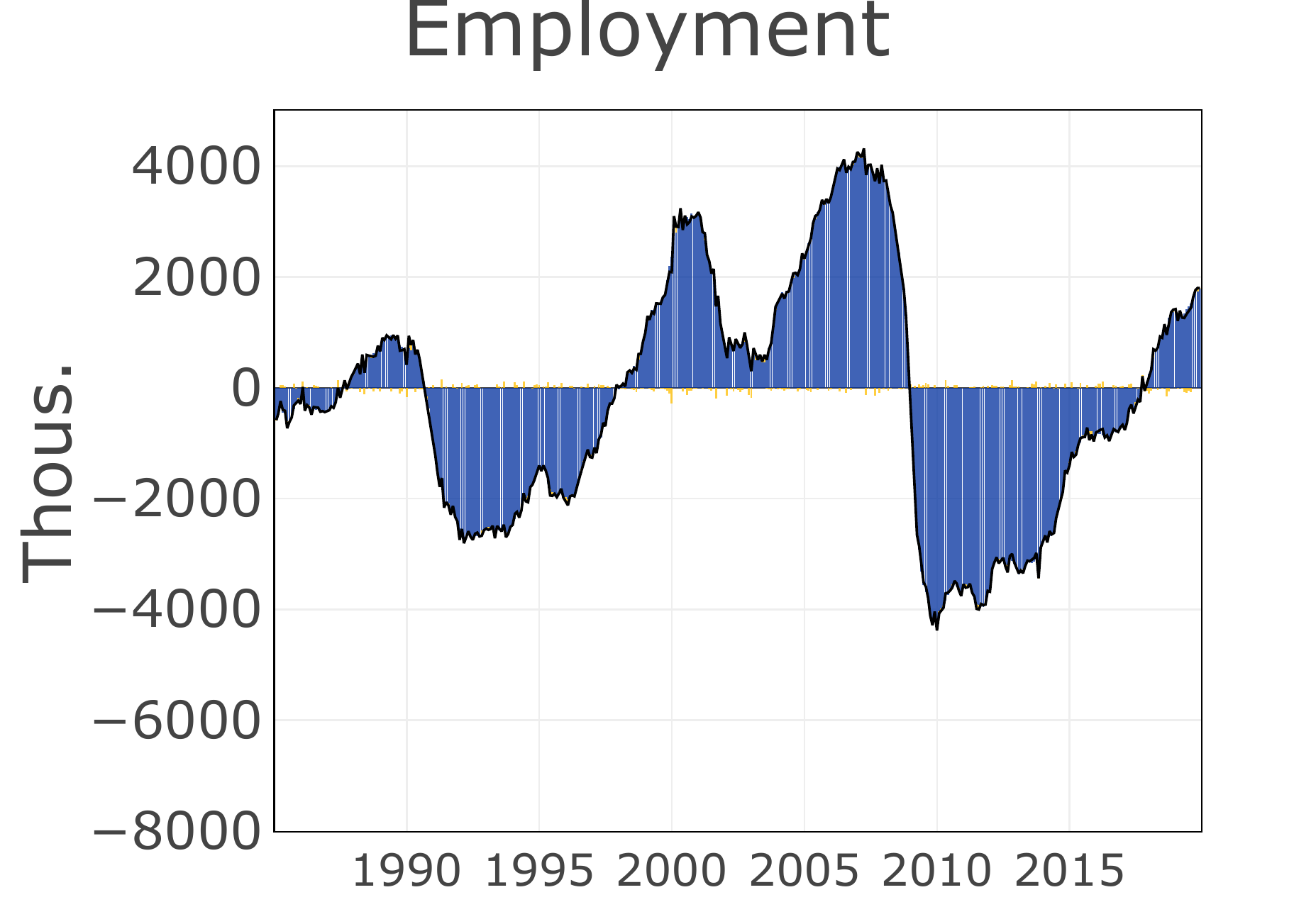}
    \end{subfigure}

    \begin{subfigure}[b]{0.24\textwidth}
        \centering
        \includegraphics[width=\textwidth]{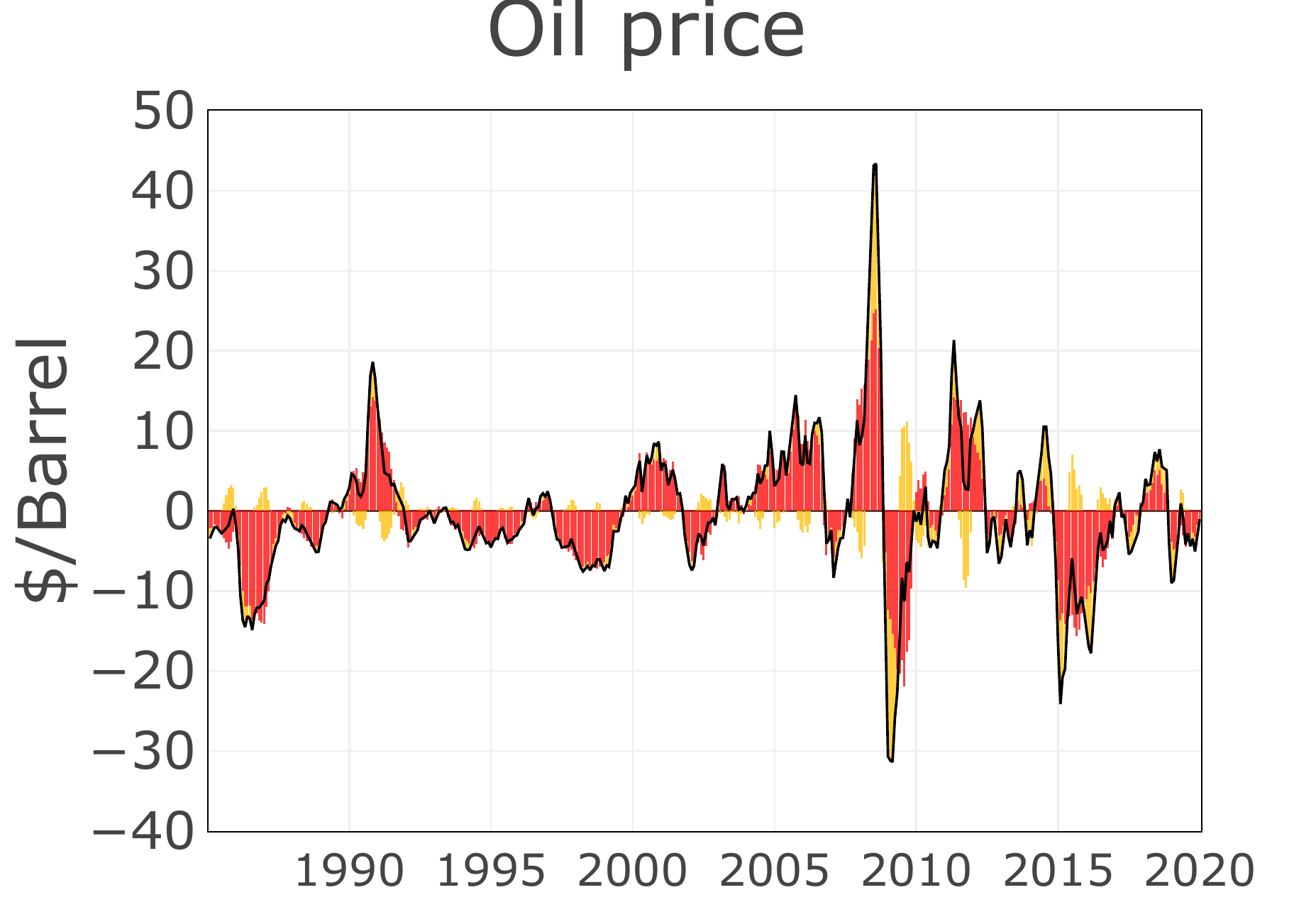}
    \end{subfigure}
    \hfill
    \begin{subfigure}[b]{0.24\textwidth}
        \centering
        \includegraphics[width=\textwidth]{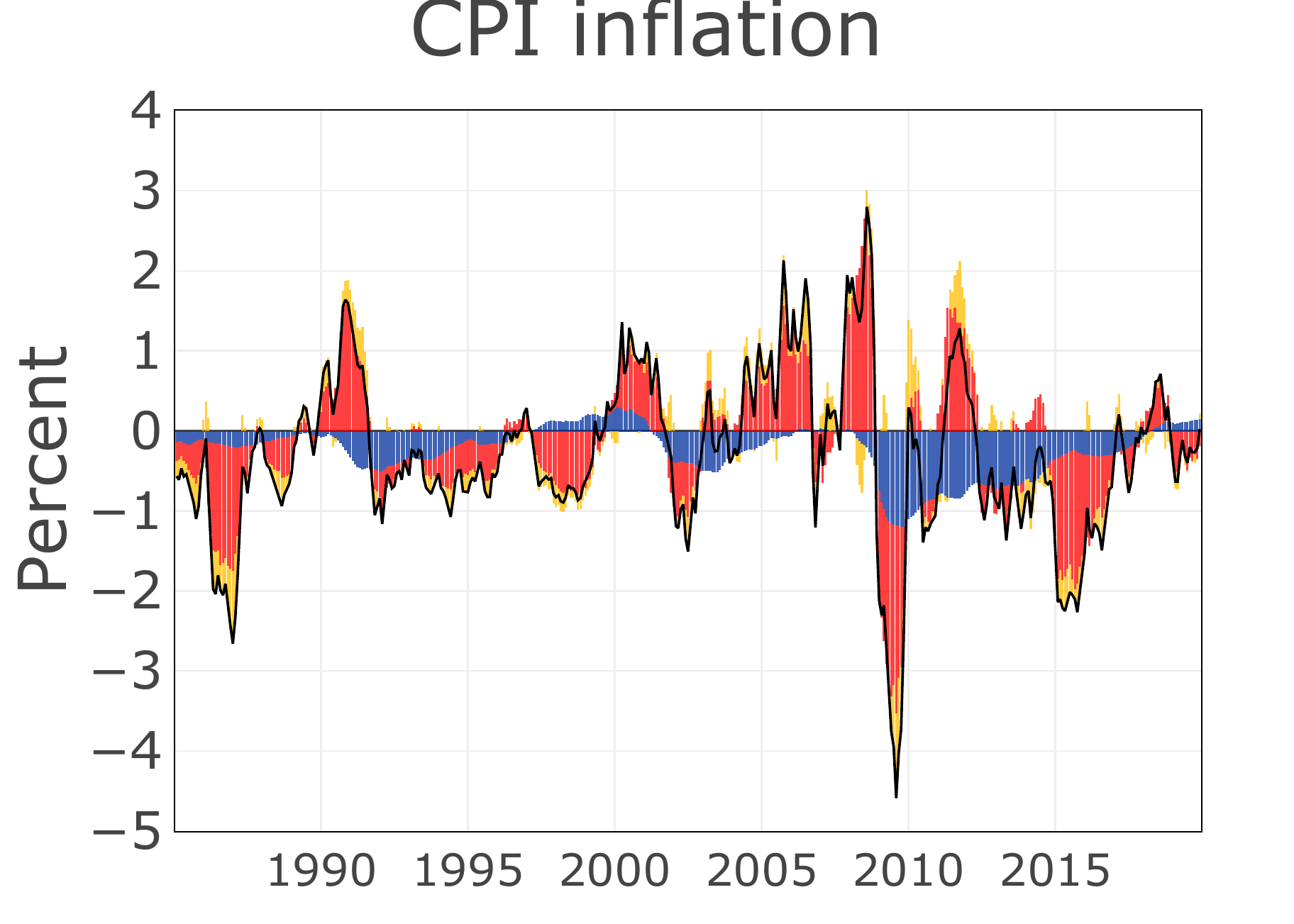}
    \end{subfigure}
    \hfill
    \vrule\
    \begin{subfigure}[b]{0.24\textwidth}
        \centering
        \includegraphics[width=\textwidth]{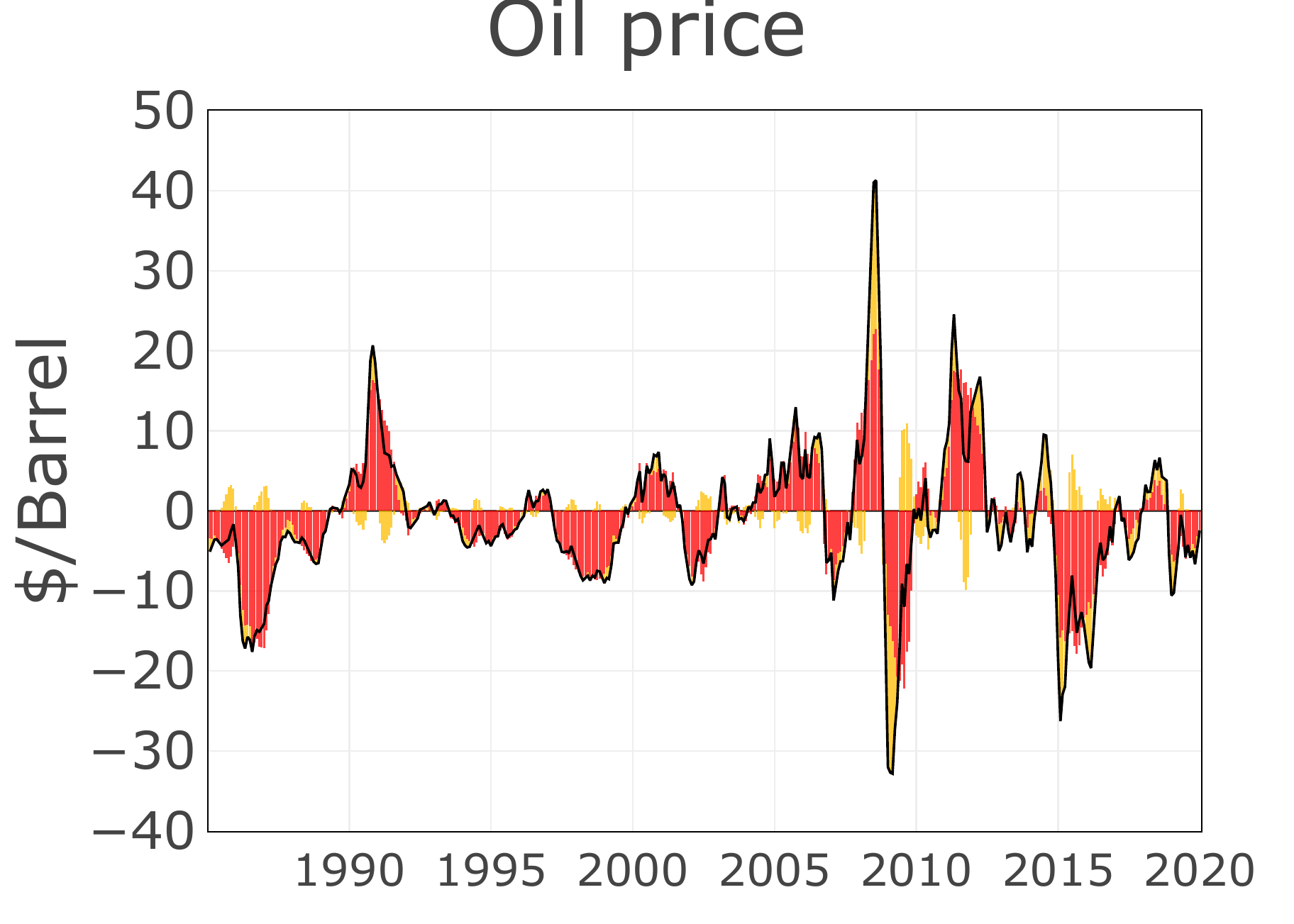}
    \end{subfigure}
	\hfill
    \begin{subfigure}[b]{0.24\textwidth}
        \centering
        \includegraphics[width=\textwidth]{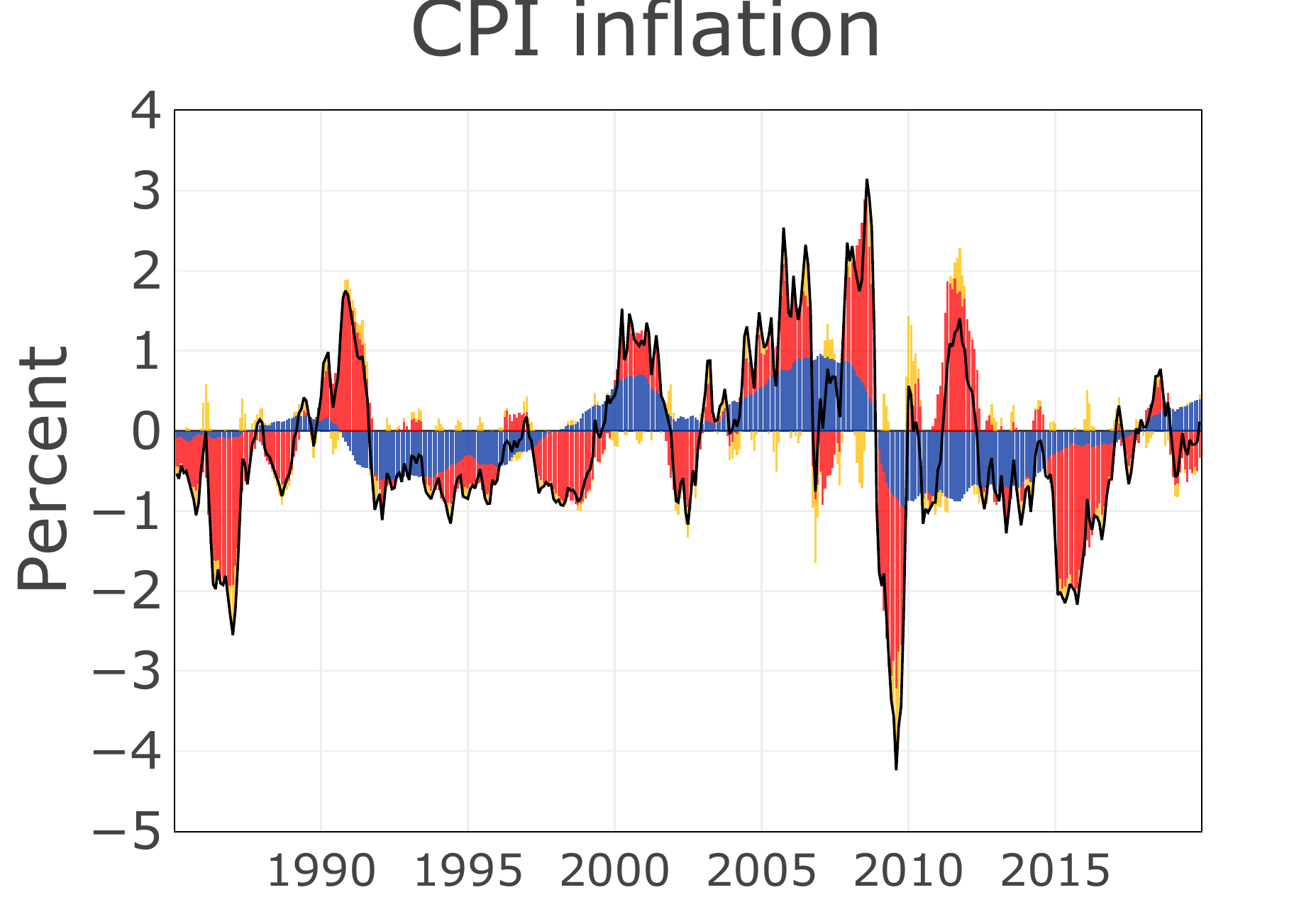}
    \end{subfigure}

    \begin{subfigure}[b]{0.24\textwidth}
        \centering
        \includegraphics[width=\textwidth]{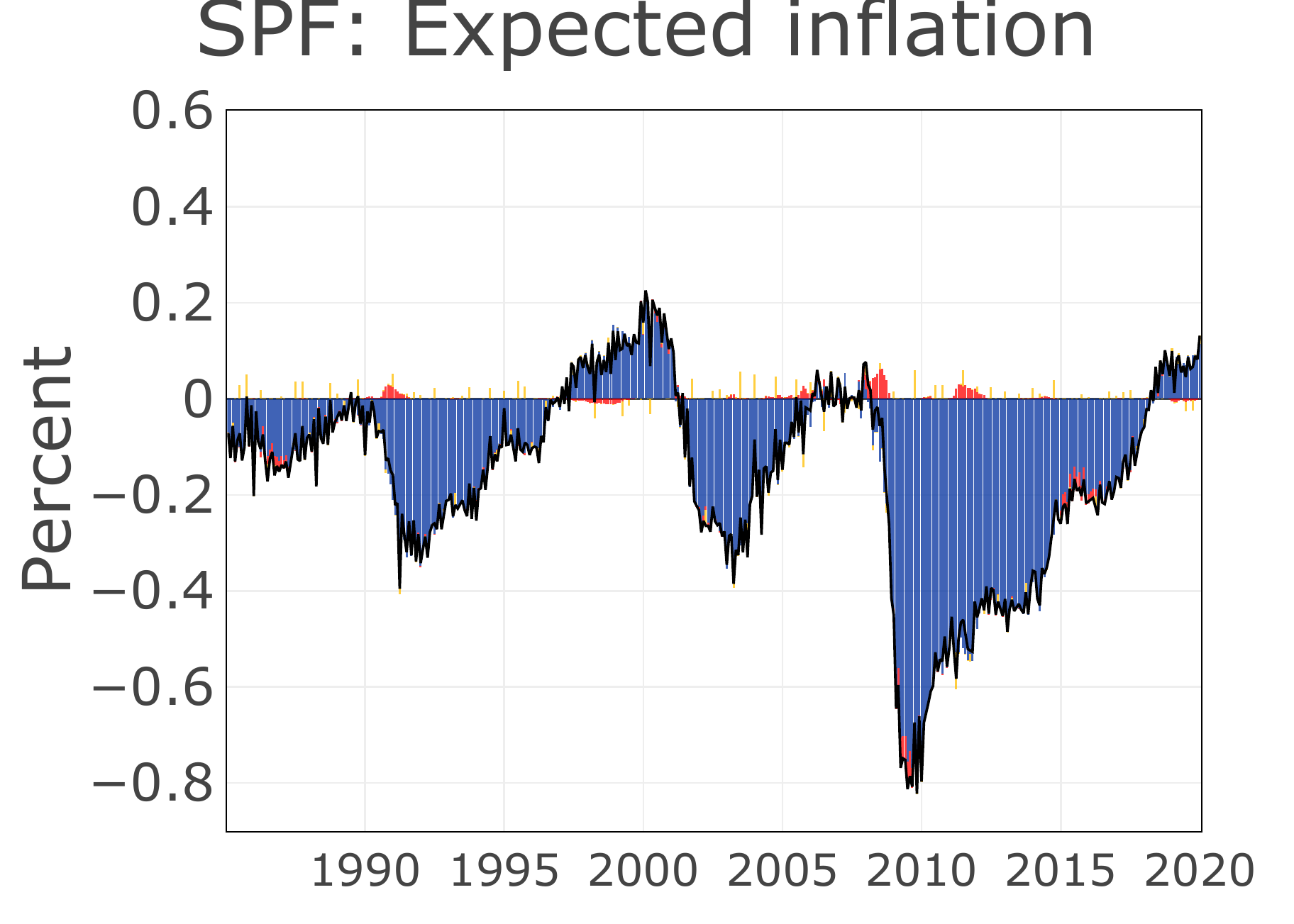}
    \end{subfigure}
    \hfill
    \begin{subfigure}[b]{0.24\textwidth}
        \centering
        \includegraphics[width=\textwidth]{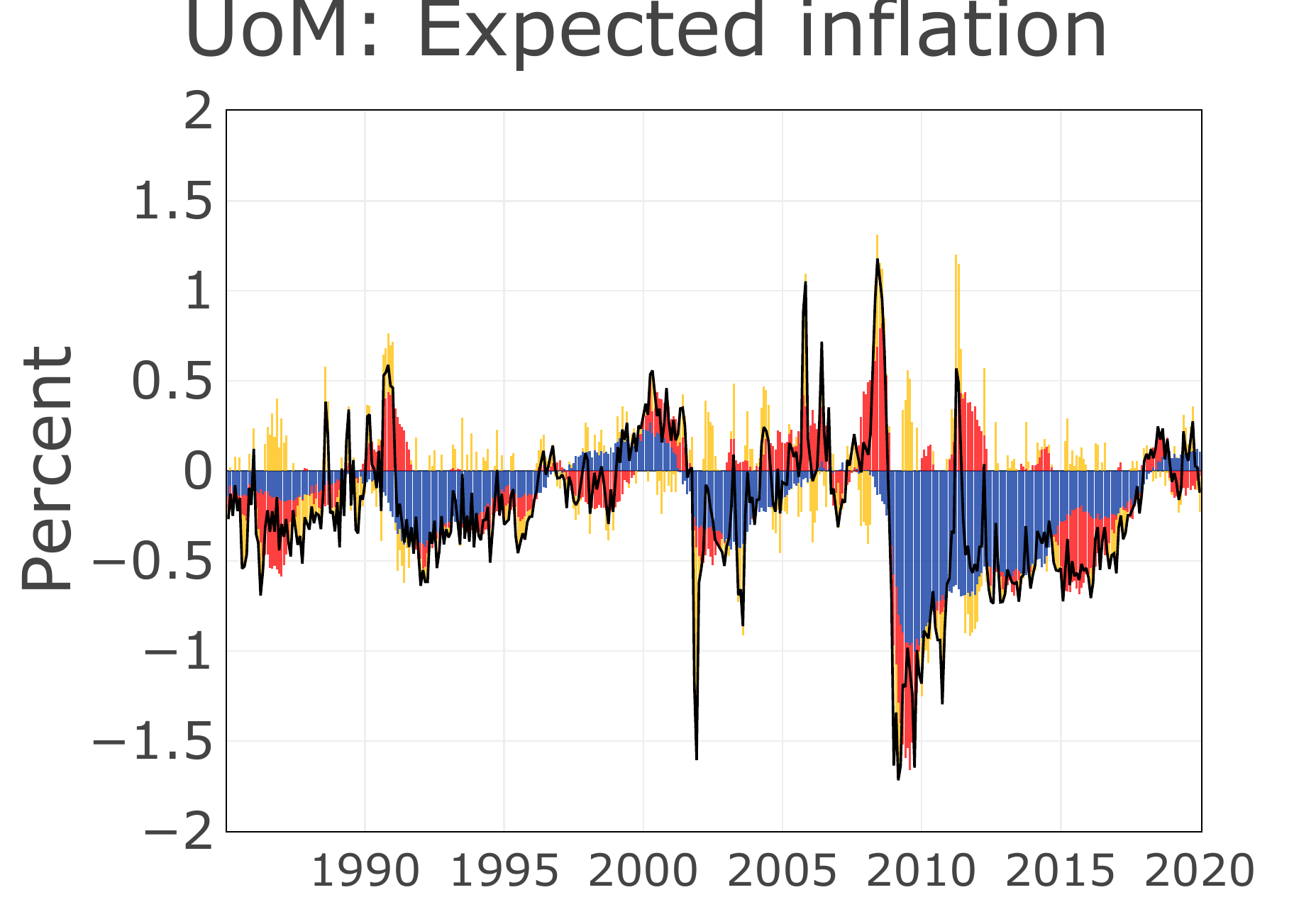}
    \end{subfigure}
    \hfill
    \vrule\
    \begin{subfigure}[b]{0.24\textwidth}
        \centering
        \includegraphics[width=\textwidth]{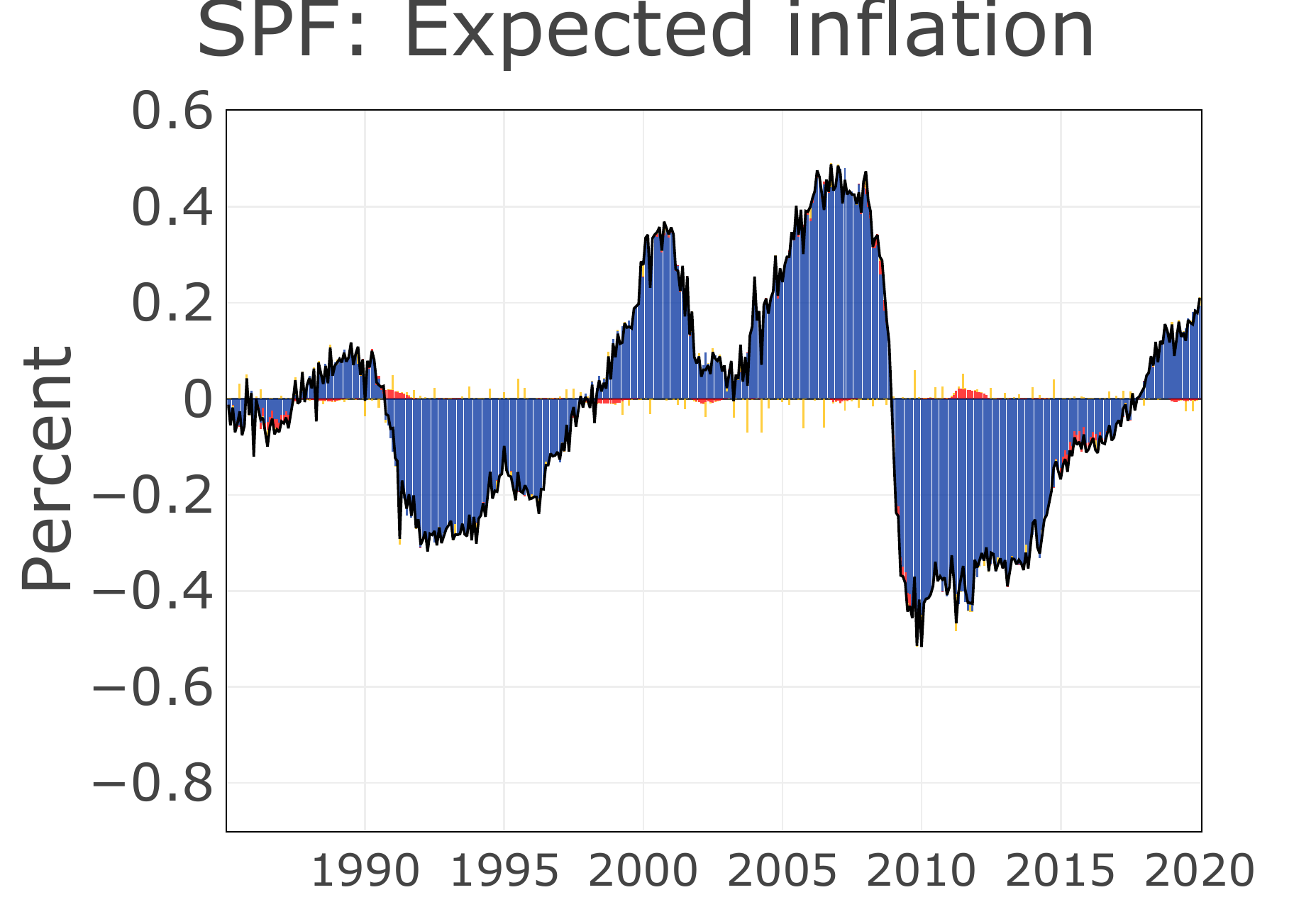}
    \end{subfigure}
	\hfill
    \begin{subfigure}[b]{0.24\textwidth}
        \centering
        \includegraphics[width=\textwidth]{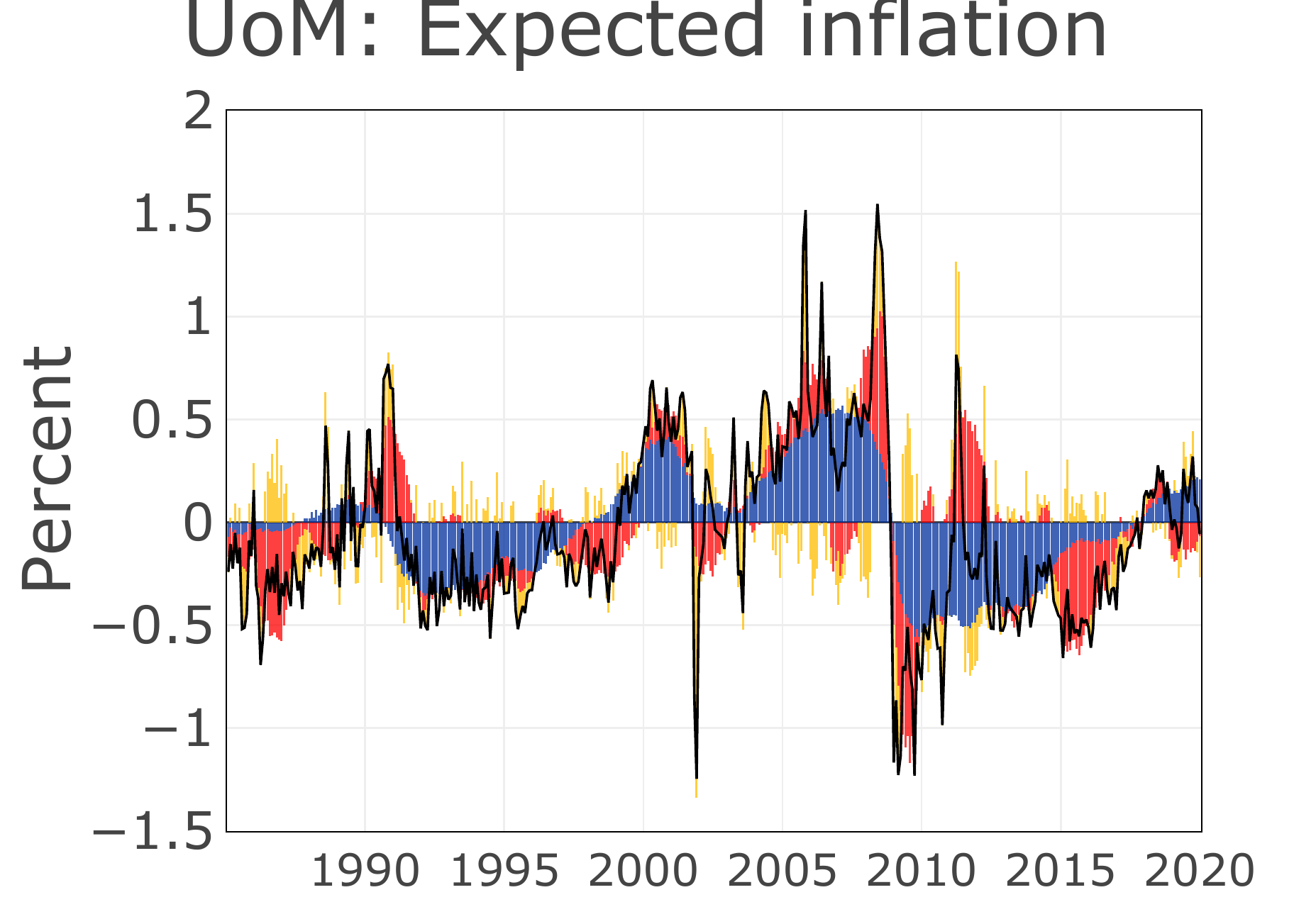}
    \end{subfigure}

    \begin{subfigure}[b]{0.24\textwidth}
        \centering
        \includegraphics[width=\textwidth]{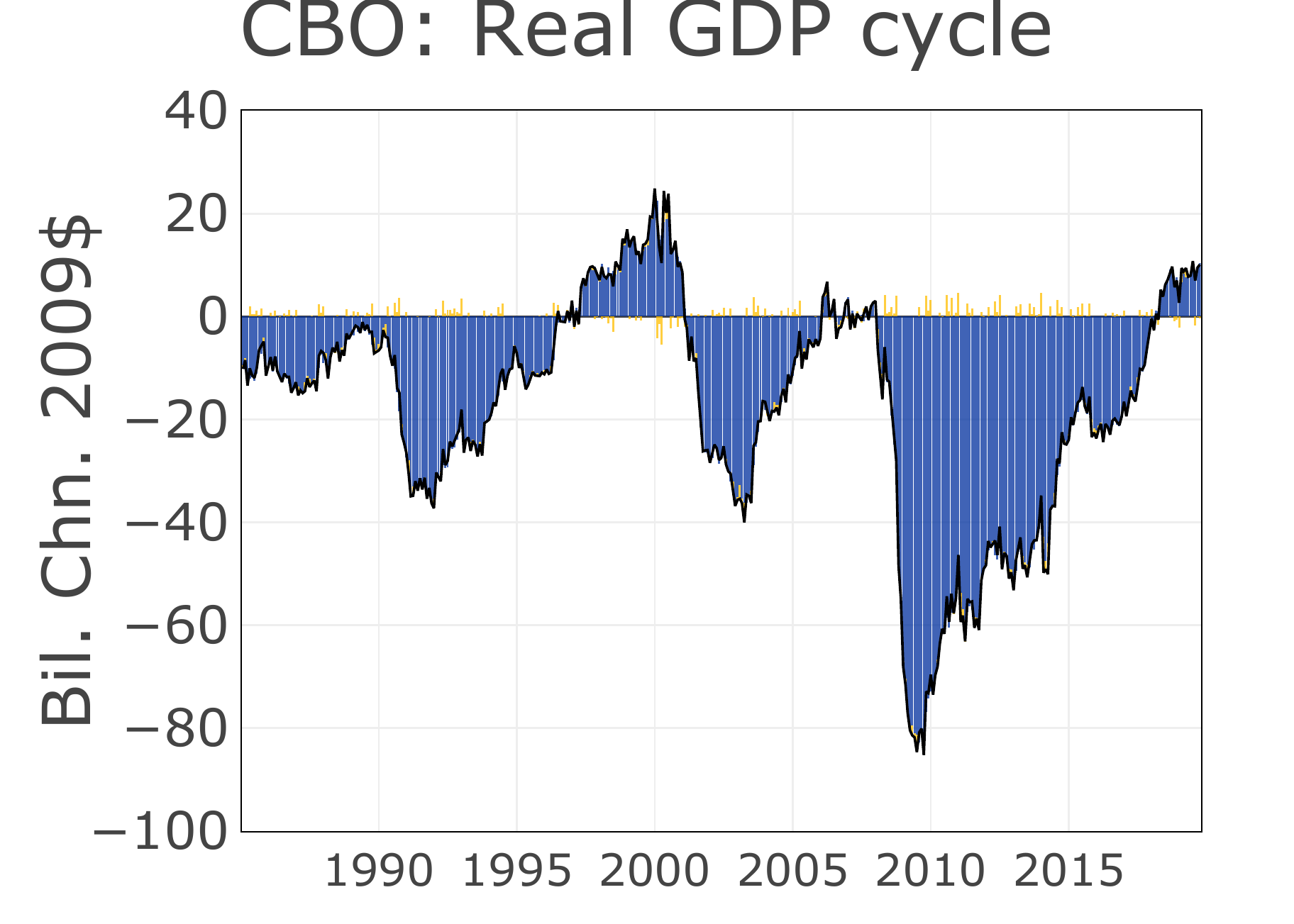}
    \end{subfigure}
	\hfill
	\hspace{0.24\textwidth}
    \vrule\
    \hspace{0.24\textwidth}
    \begin{subfigure}[b]{0.24\textwidth}
        \centering
        \includegraphics[width=\textwidth]{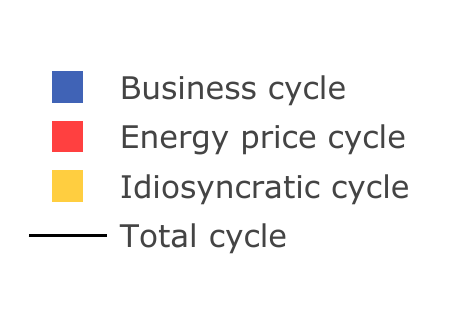}
    \end{subfigure}
	\hfill
    \caption{Historical decomposition of the stationary components of all the variables in the tracking model (left) and the undisciplined model (right).}
    \label{fig:historical_reconstruction}
\end{figure}

The two models provide a similar overall reading of developments in the US economy, with some key differences. \autoref{fig:historical_reconstruction} compares the two sets of results and reports the cyclical components of all variables for the two models: (i) the output gap and the business cycle (blue), (ii) the energy price component (red), and a residual idiosyncratic component reflecting measurement error and unmodelled components. 

The tracking model -- in line with the assessment of the CBO onto which it is geared -- estimates an output gap that shows significant contractions in the cyclical component of real GDP after the early 1990s recession -- i.e., the dot-com bubble and the Great Recession (see \autoref{fig:historical_reconstruction}). This contrasts with the assessment of the undisciplined model, which shows larger expansions in the late 1990s and early 2000s that culminated in the dot-com crisis and the Great Recession. The idiosyncratic component plays a small role in both specifications, which implies that differences across models in assessing the cyclical component reflect the different evaluations of potential output. Differences in output gap estimates are reflected in the cycles of employment and unemployment, which are linked to the output gap by Okun's law and in the Phillips curve part of the inflation cycle. For example, the tracking model reads negative or neutral inflation pressure from the real economy before the financial crisis while the undisciplined model identifies positive pressures.

The correlation between the unemployment  and the inflation cycles is similar across models, -0.41 for the undisciplined and -0.36 for the tracking. However, the common cycle between inflation and the real economy is masked by the highly volatile energy component unrelated to domestic business cycle fluctuations. Consequently, the overall cyclical part of inflation is similar across models.

\begin{figure}[t!]
    \centering
	{\bf \footnotesize Tracking Model} \hspace{0.24\textwidth} {\bf \footnotesize  Undisciplined Model}
	\vfill
    \begin{subfigure}[b]{0.24\textwidth}
        \centering
        \includegraphics[width=\textwidth]{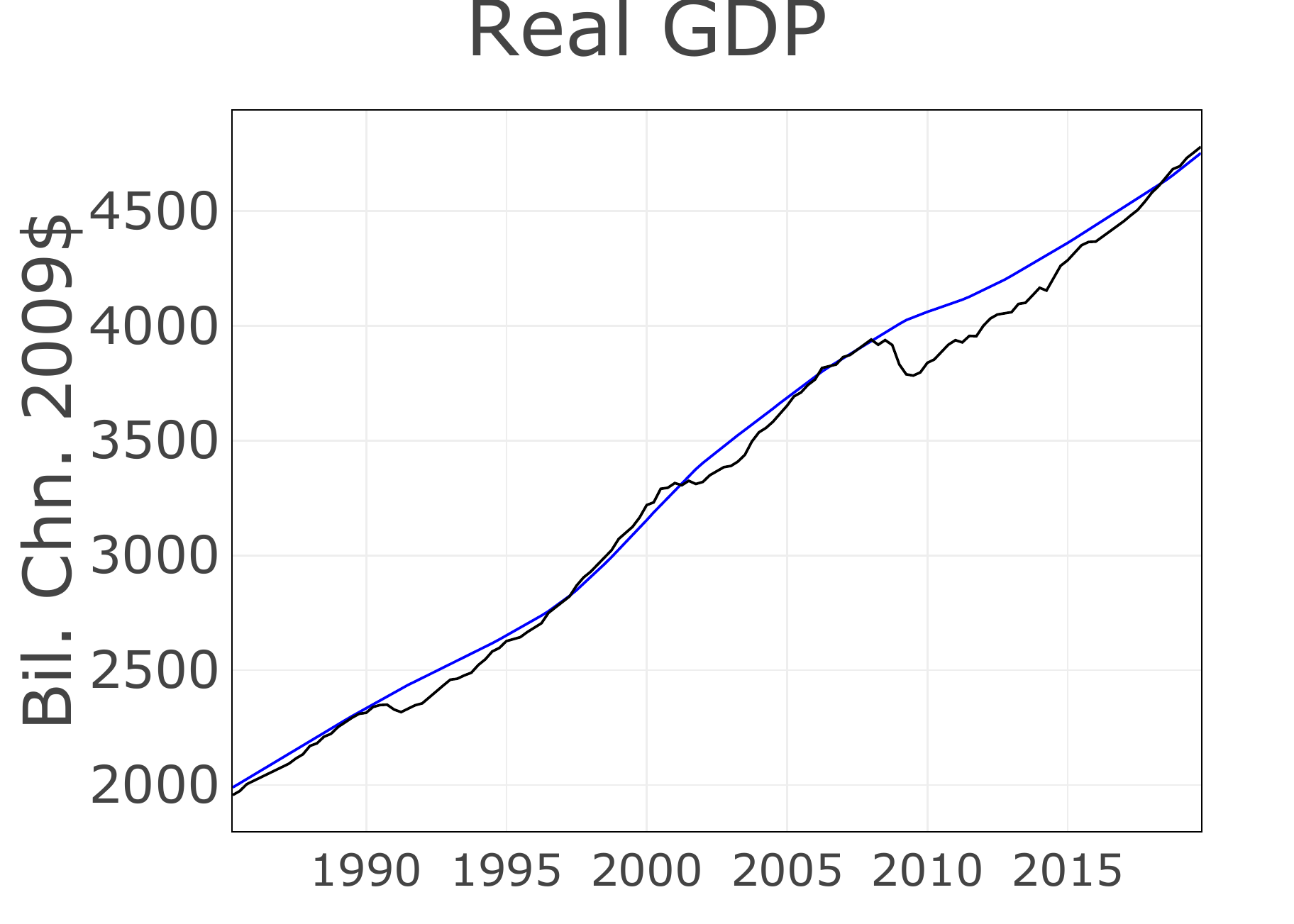}
    \end{subfigure}
    \hfill
    \begin{subfigure}[b]{0.24\textwidth}
        \centering
        \includegraphics[width=\textwidth]{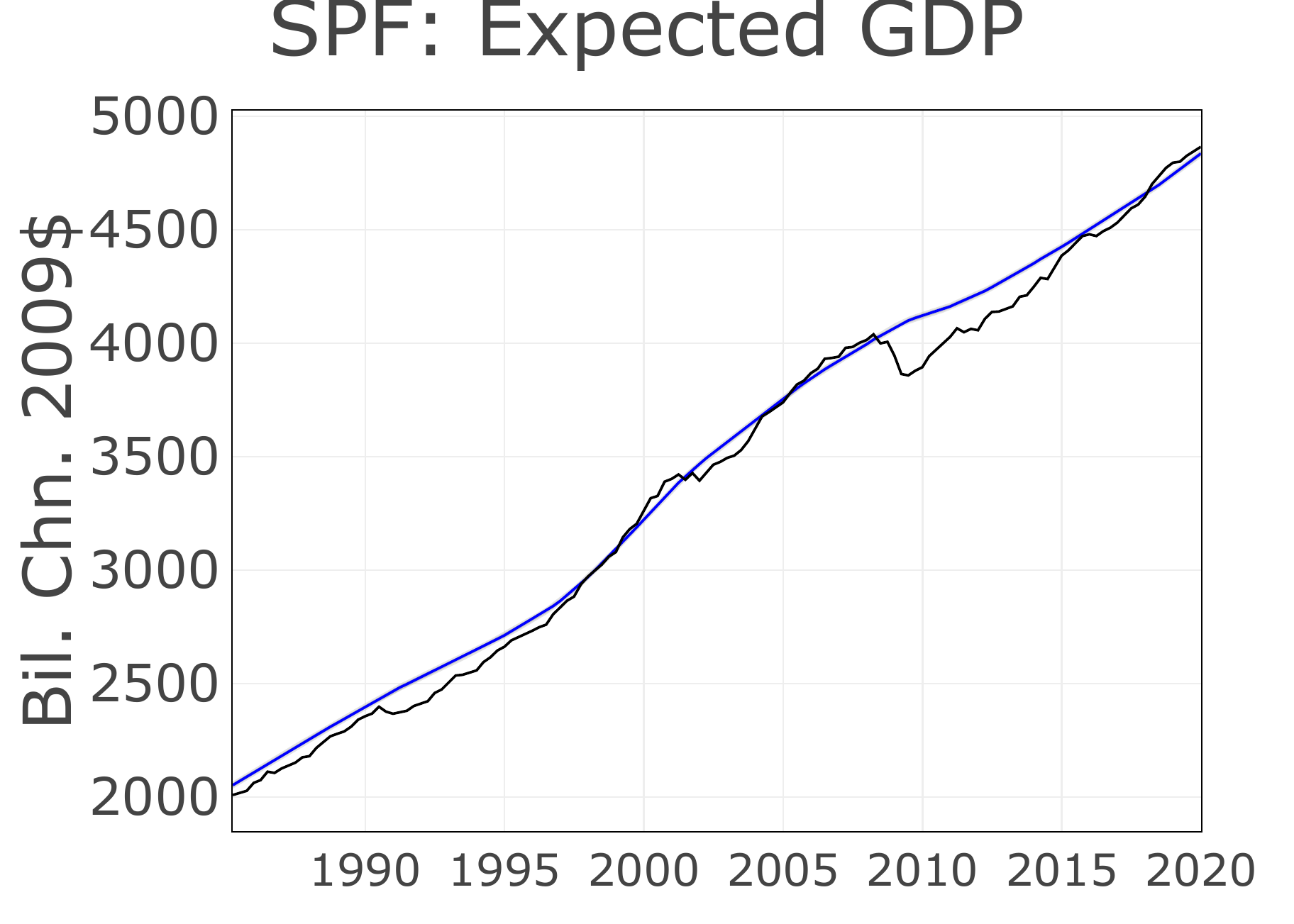}
    \end{subfigure}
    \hfill
    \vrule\
    \begin{subfigure}[b]{0.24\textwidth}
        \centering
        \includegraphics[width=\textwidth]{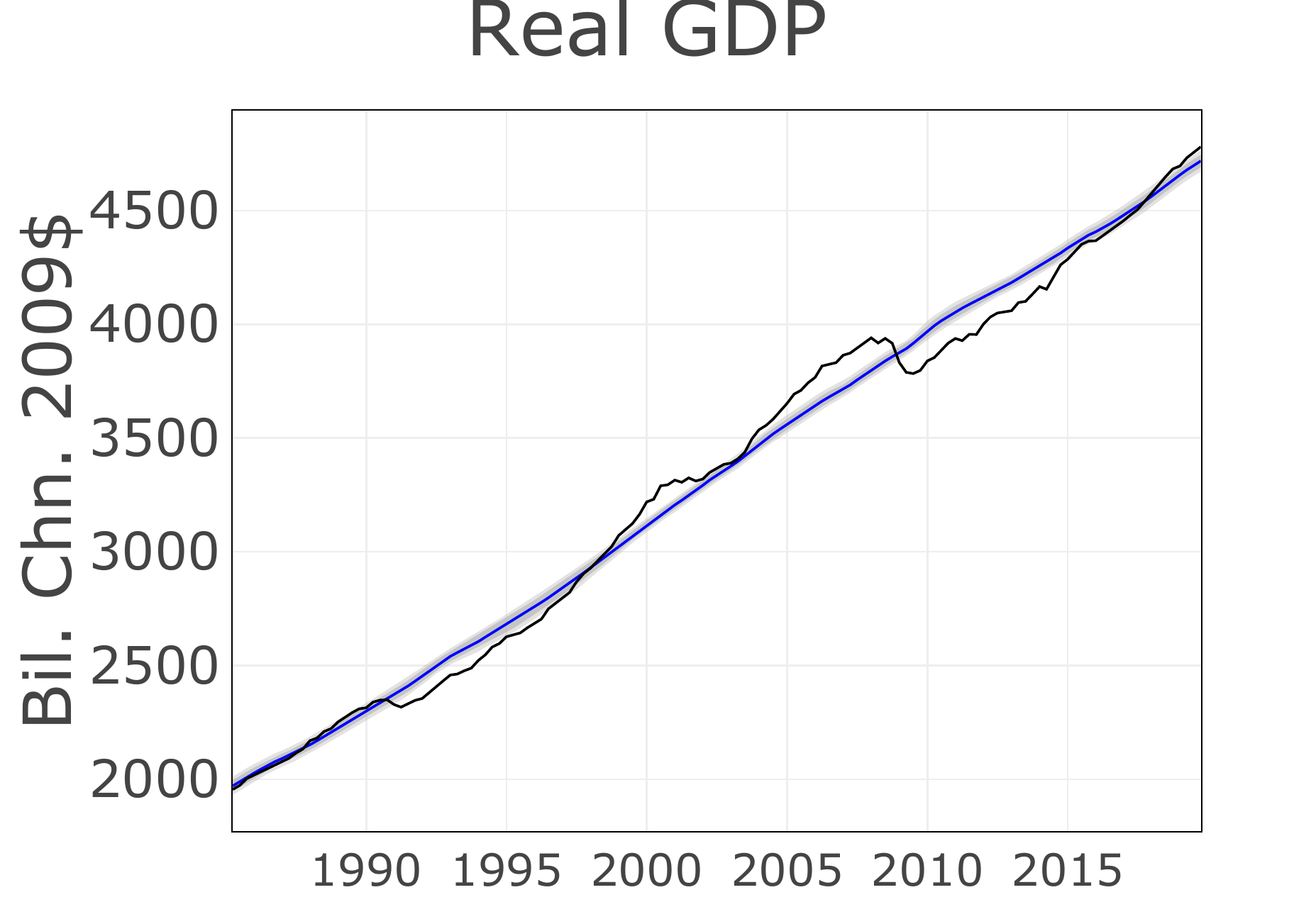}
    \end{subfigure}
	\hfill
    \begin{subfigure}[b]{0.24\textwidth}
        \centering
        \includegraphics[width=\textwidth]{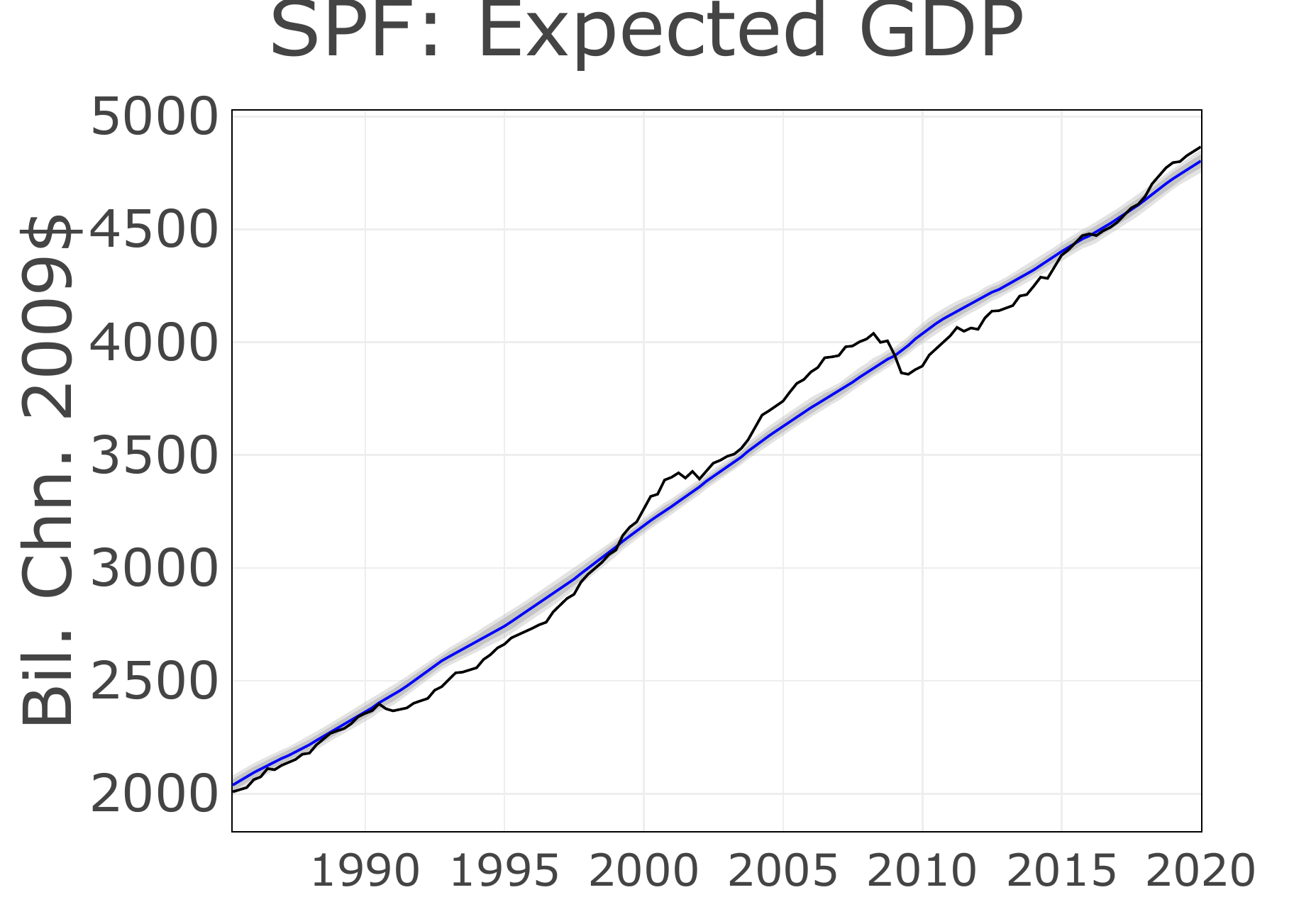}
    \end{subfigure}

    \begin{subfigure}[b]{0.24\textwidth}
        \centering
        \includegraphics[width=\textwidth]{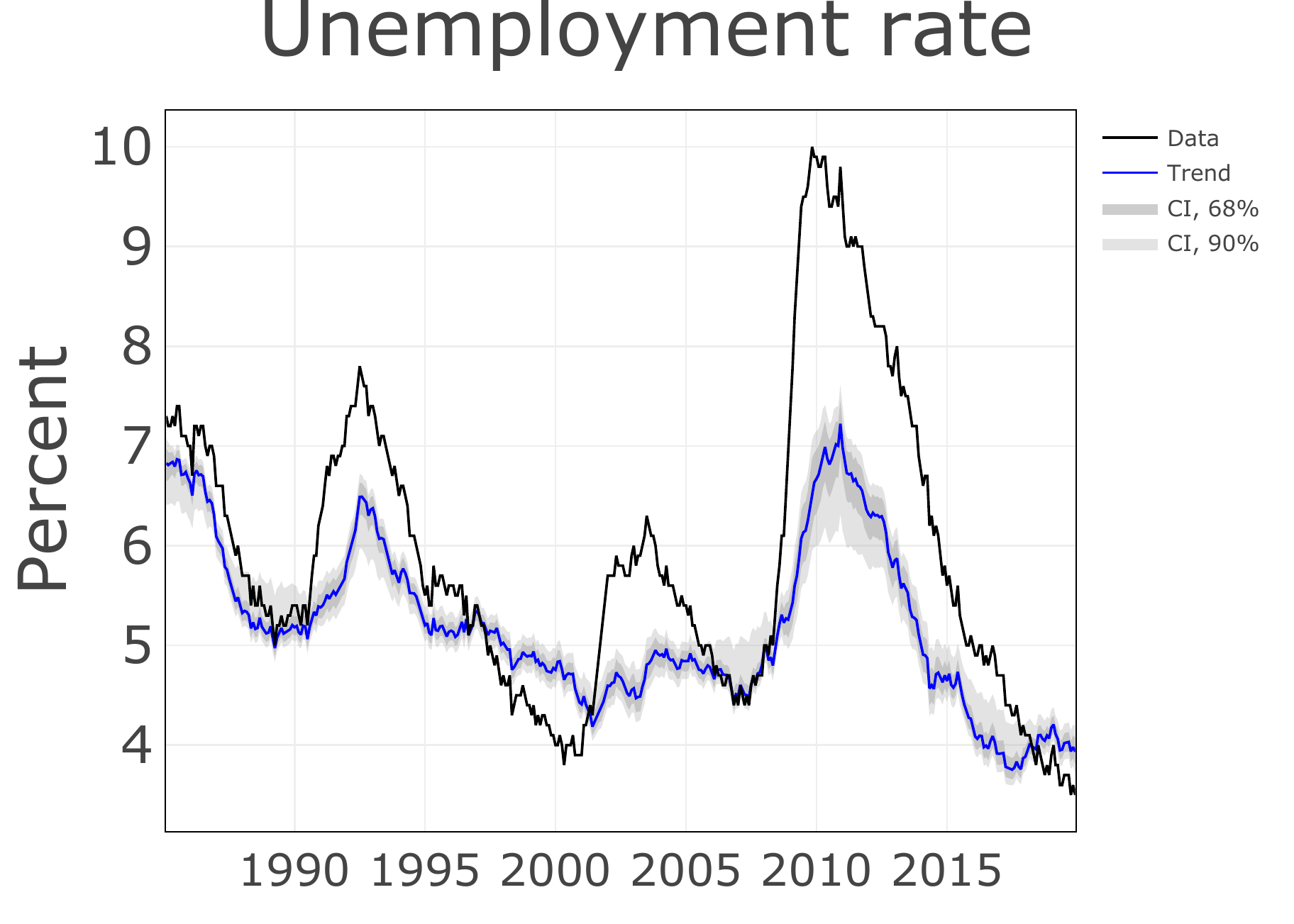}
    \end{subfigure}
    \hfill
    \begin{subfigure}[b]{0.24\textwidth}
        \centering
        \includegraphics[width=\textwidth]{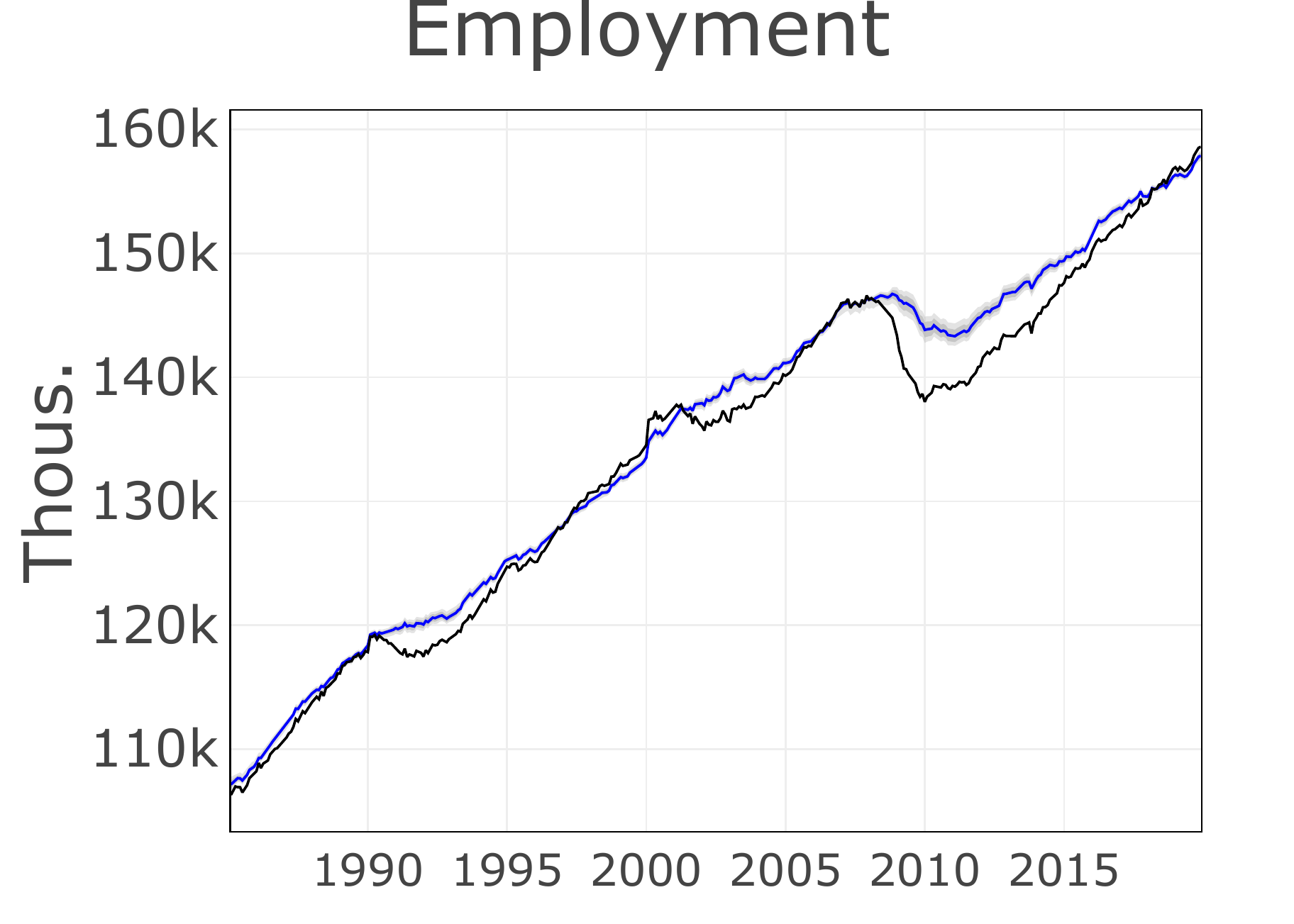}
    \end{subfigure}
    \hfill
    \vrule\
    \begin{subfigure}[b]{0.24\textwidth}
        \centering
        \includegraphics[width=\textwidth]{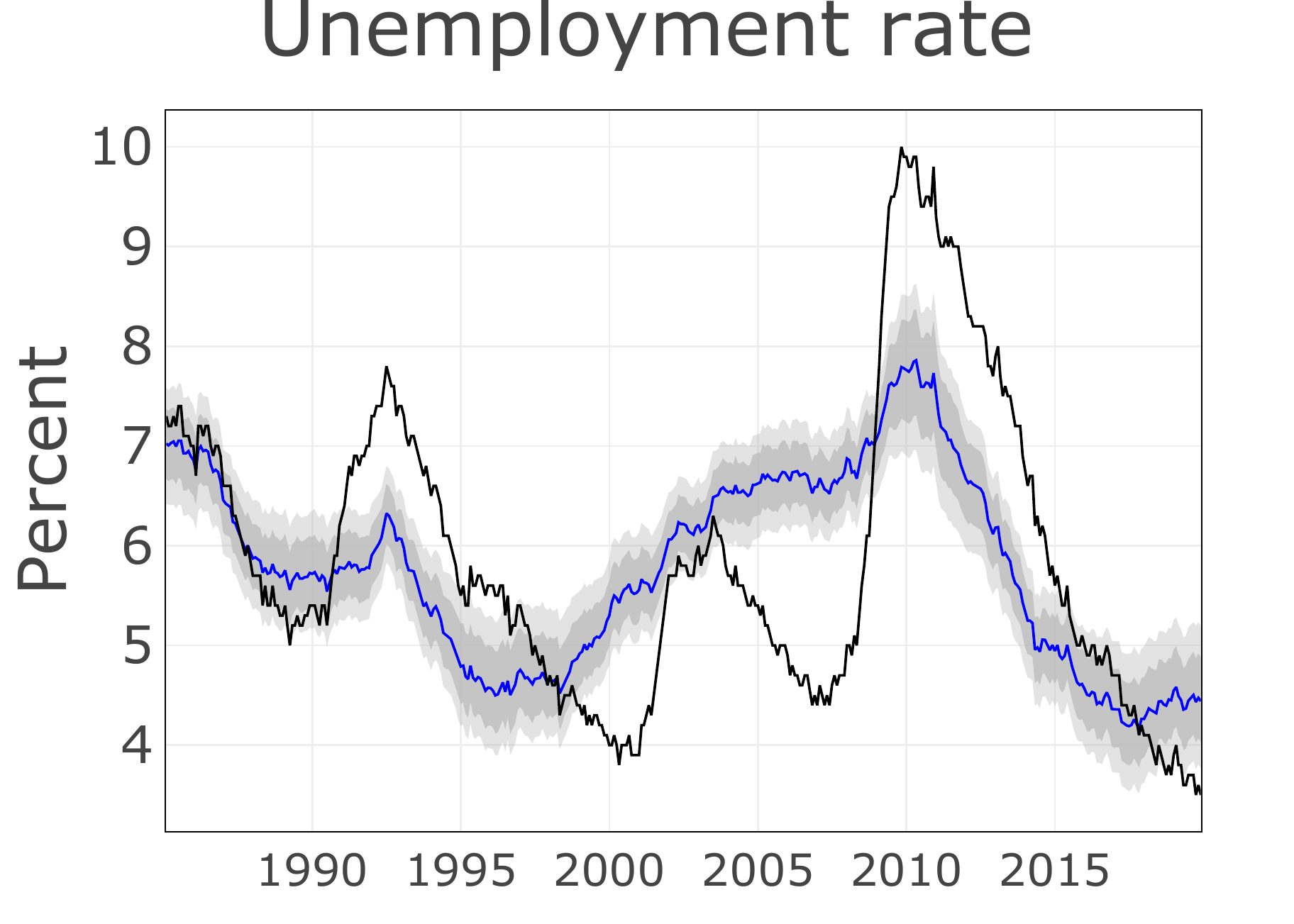}
    \end{subfigure}
	\hfill
    \begin{subfigure}[b]{0.24\textwidth}
        \centering
        \includegraphics[width=\textwidth]{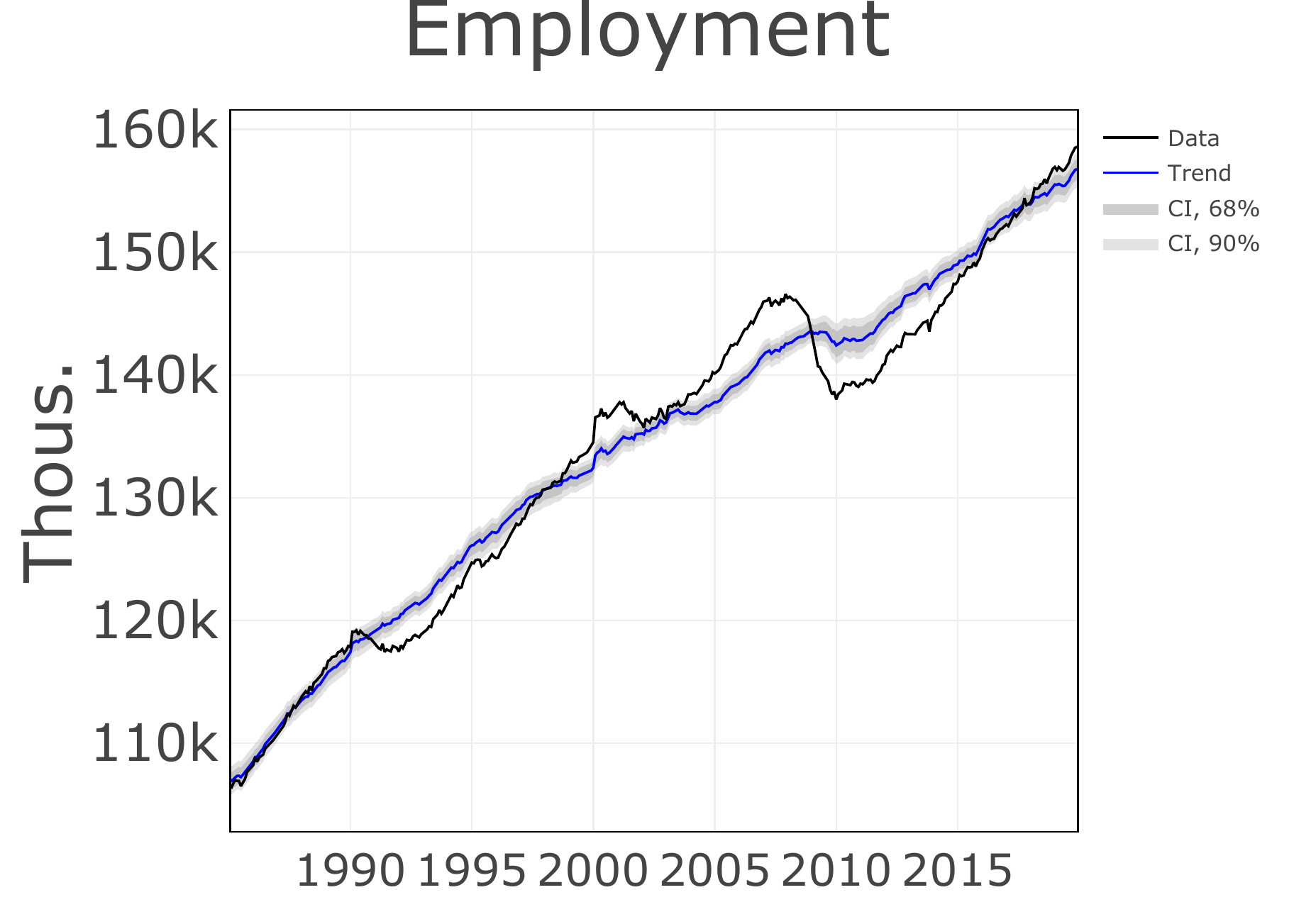}
    \end{subfigure}

    \begin{subfigure}[b]{0.24\textwidth}
        \centering
        \includegraphics[width=\textwidth]{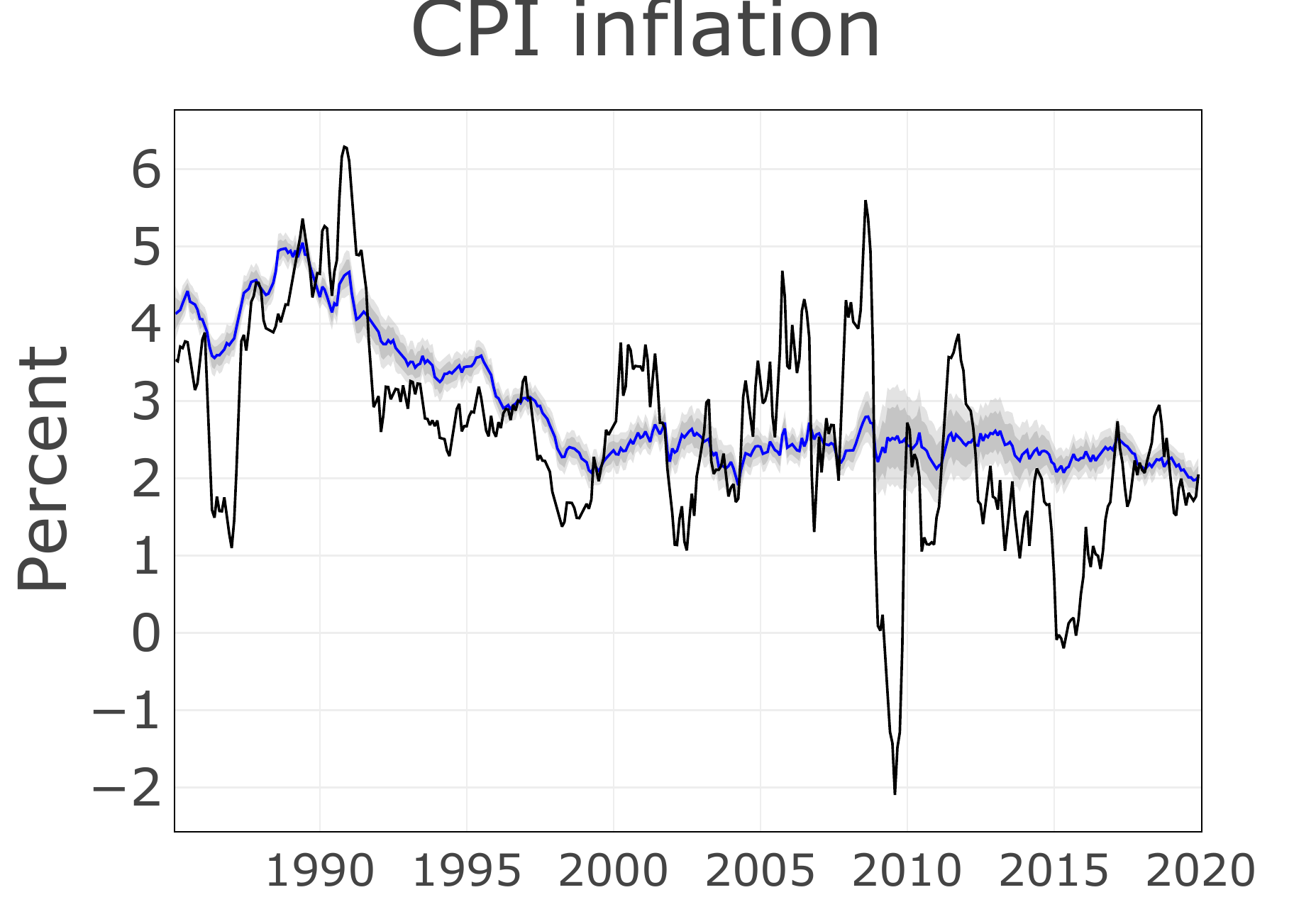}
    \end{subfigure}
    \hfill
    \begin{subfigure}[b]{0.24\textwidth}
        \centering
        \includegraphics[width=\textwidth]{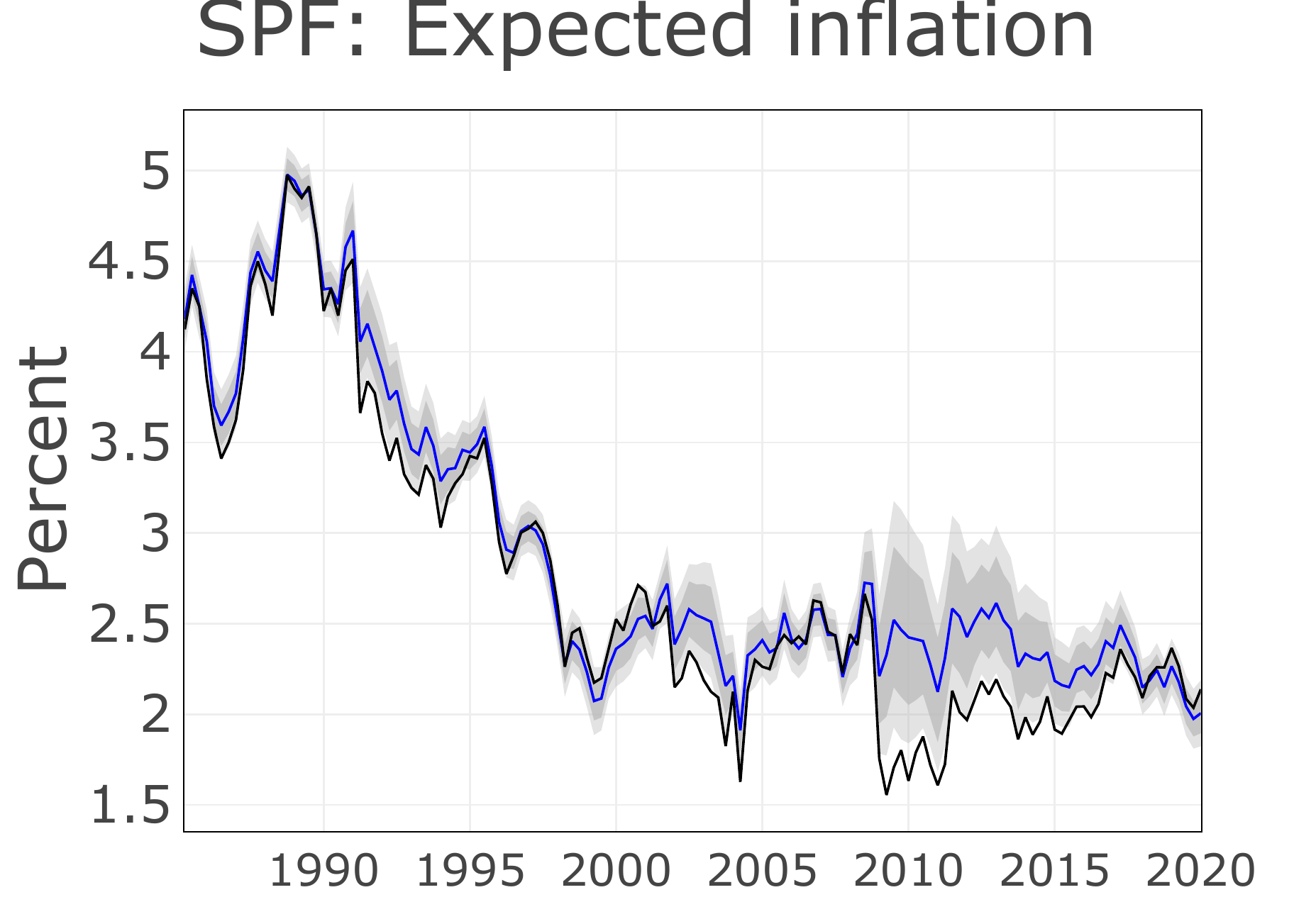}
    \end{subfigure}
    \hfill
    \vrule\
    \begin{subfigure}[b]{0.24\textwidth}
        \centering
        \includegraphics[width=\textwidth]{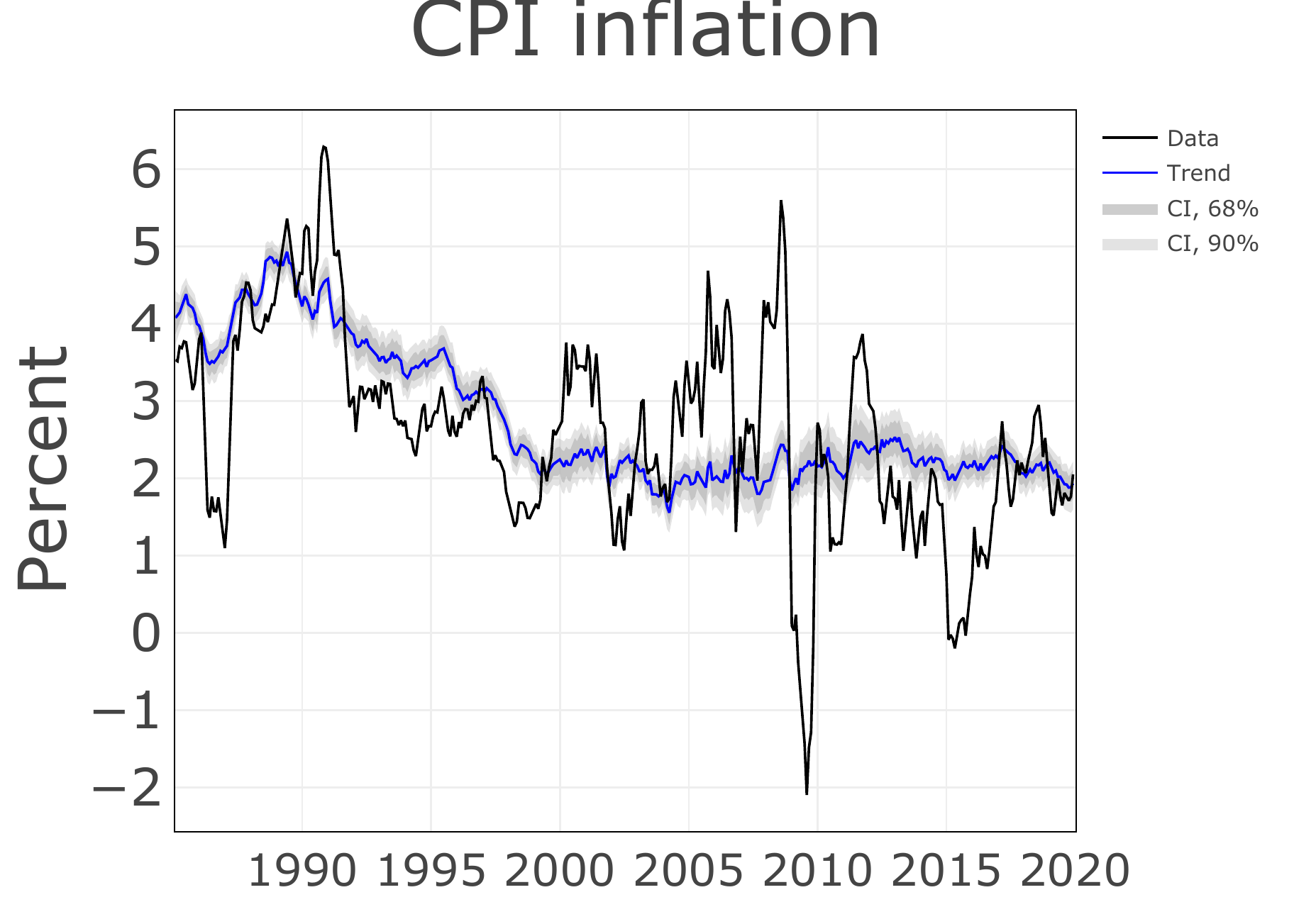}
    \end{subfigure}
	\hfill
    \begin{subfigure}[b]{0.24\textwidth}
        \centering
        \includegraphics[width=\textwidth]{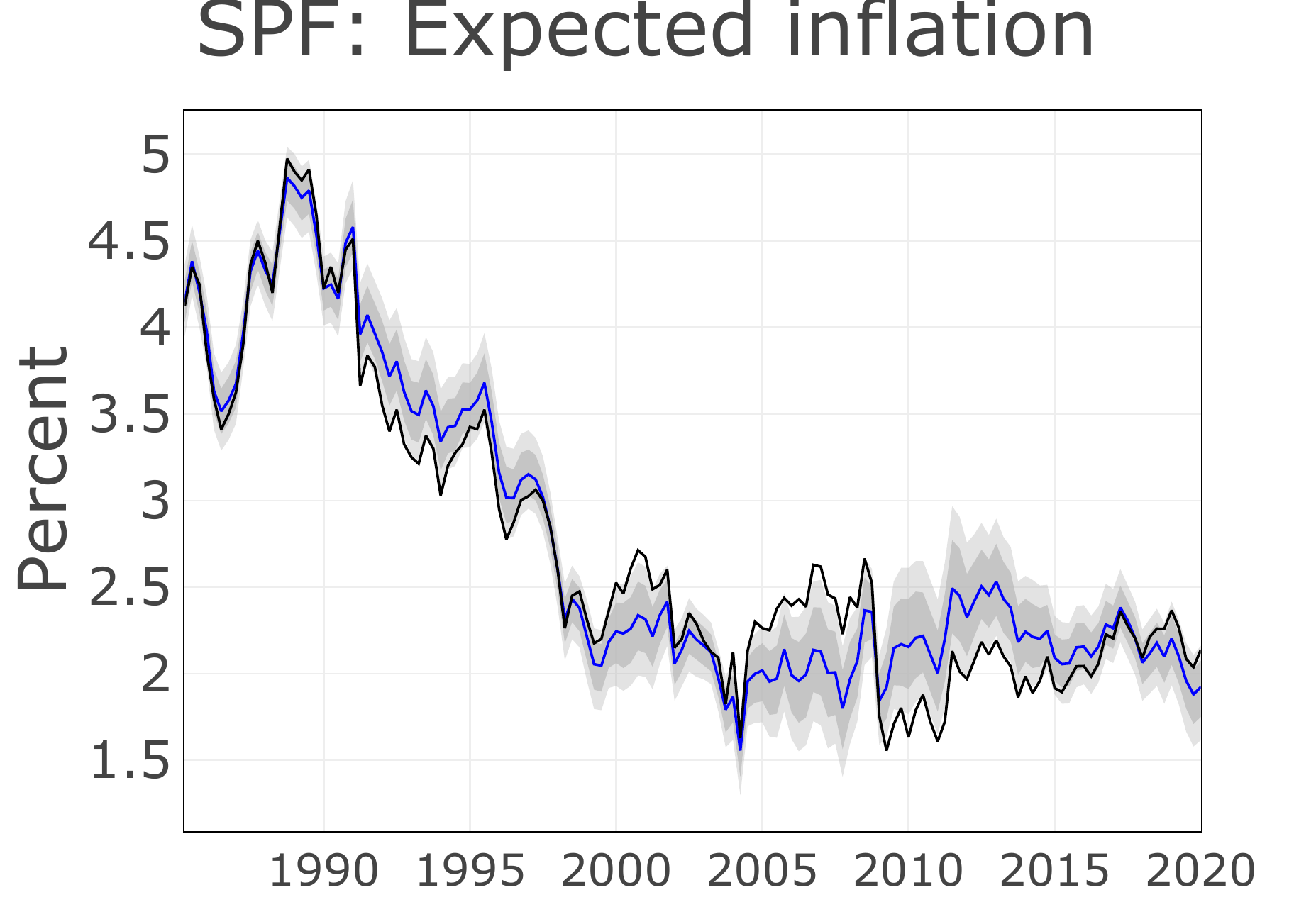}
    \end{subfigure}

    \begin{subfigure}[b]{0.24\textwidth}
        \centering
        \includegraphics[width=\textwidth]{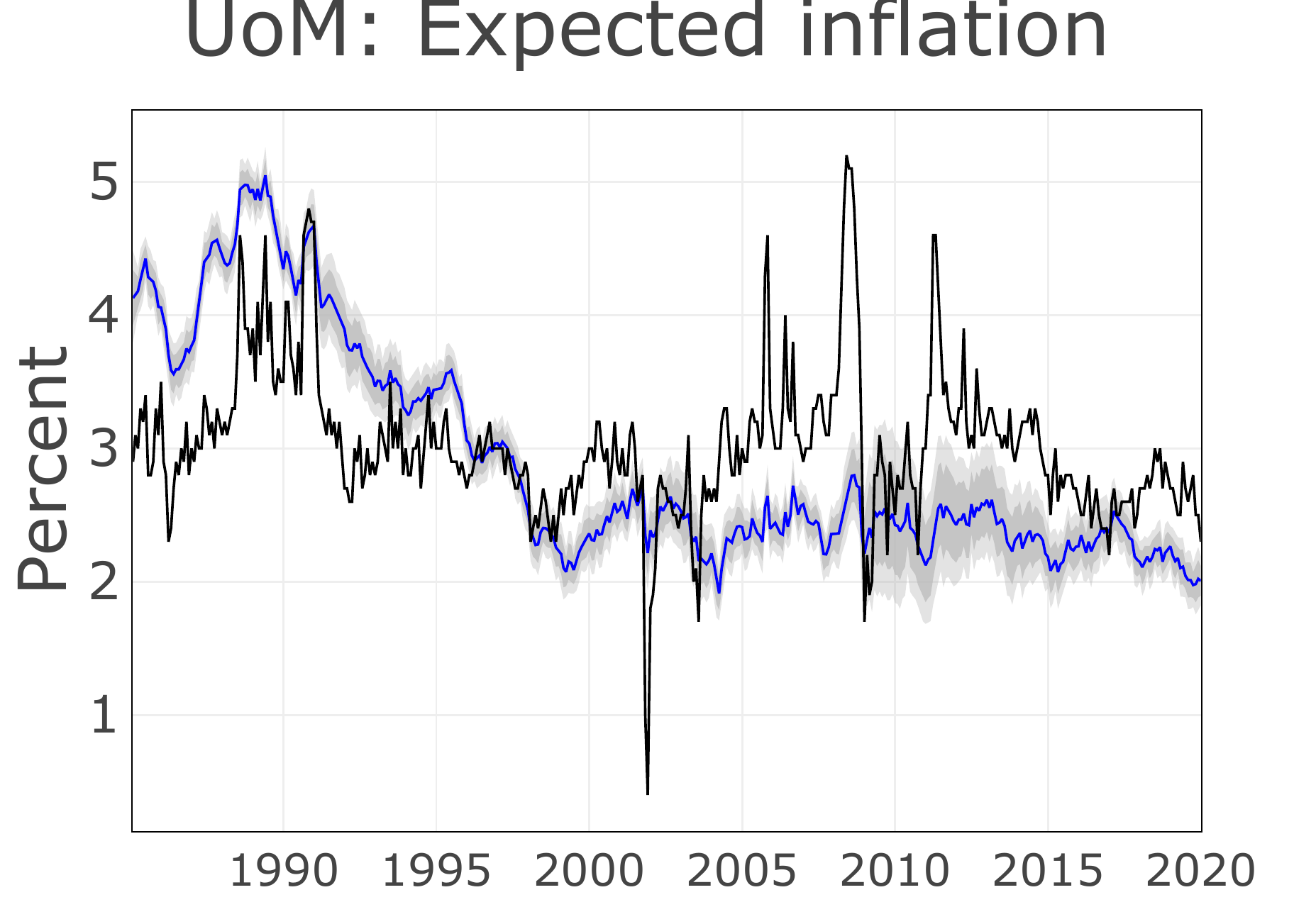}
    \end{subfigure}
    \hfill
	\hspace{0.24\textwidth}
	\hfill
	\vrule\
    \begin{subfigure}[b]{0.24\textwidth}
        \centering
        \includegraphics[width=\textwidth]{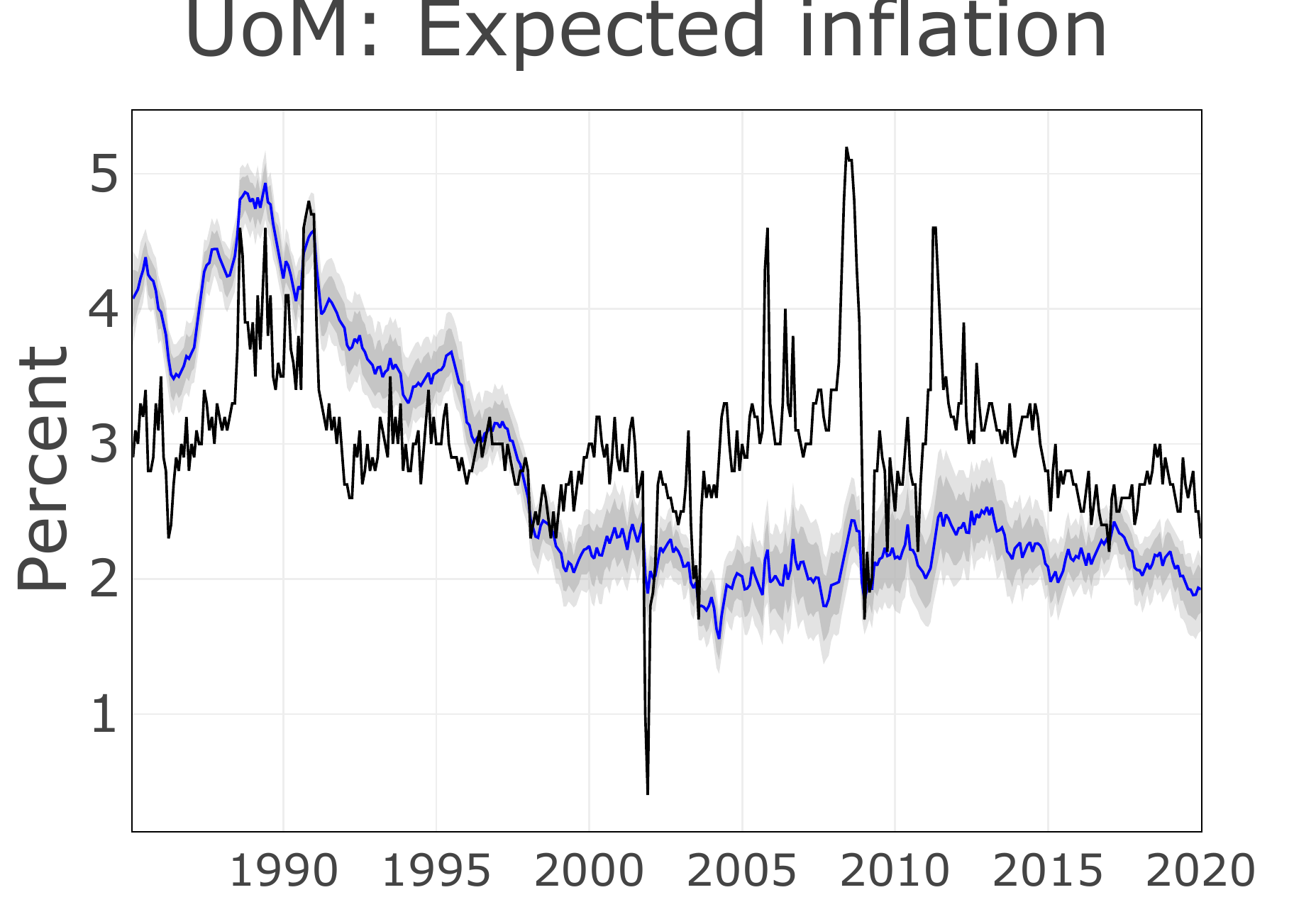}
    \end{subfigure}
    \hfill
    \begin{subfigure}[b]{0.24\textwidth}
        \centering
        \includegraphics[width=\textwidth]{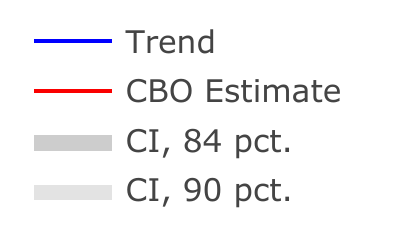}
    \end{subfigure}

    \caption{Trends for all the variables in the tracking model (left) and the undisciplined model (right), along with 84\% 90\% posterior coverage bands and  The charts also report the CBO estimates for output potential and NAIRU. }
    \label{fig:trends}
\end{figure}

As observed, the differences in the measure of the output gap between models are due to differences in the estimated potential, being the gap the difference between output and its trend. Estimates of the trends for all the variables are reported in \autoref{fig:trends}, where they are plotted against their associated observable variables.  It can be easily seen that the undisciplined model fits an output gap that fluctuates almost symmetrically around the trend, as would be the case in a standard Neoclassical or New Keynesian macroeconomic model. Conversely, in the model informed by the CBO, potential output is above GDP most of the time, and recessions appear as shortfalls against this higher level. The undisciplined model attributes a larger part of the output variance to the trend, interpreting the slowdown since 2001, especially since 2008, as a change to potential output rather than as cyclical fluctuations. Consequently, the estimate of the NAIRU in the second half of the sample is higher. Not surprisingly, trend inflation is almost identical across the two models.

\section{A real-time evaluation in times of pandemics}\label{sec:real-time} 

\begin{figure}[ht!]
	\centering
	\includegraphics[width=0.49\textwidth]{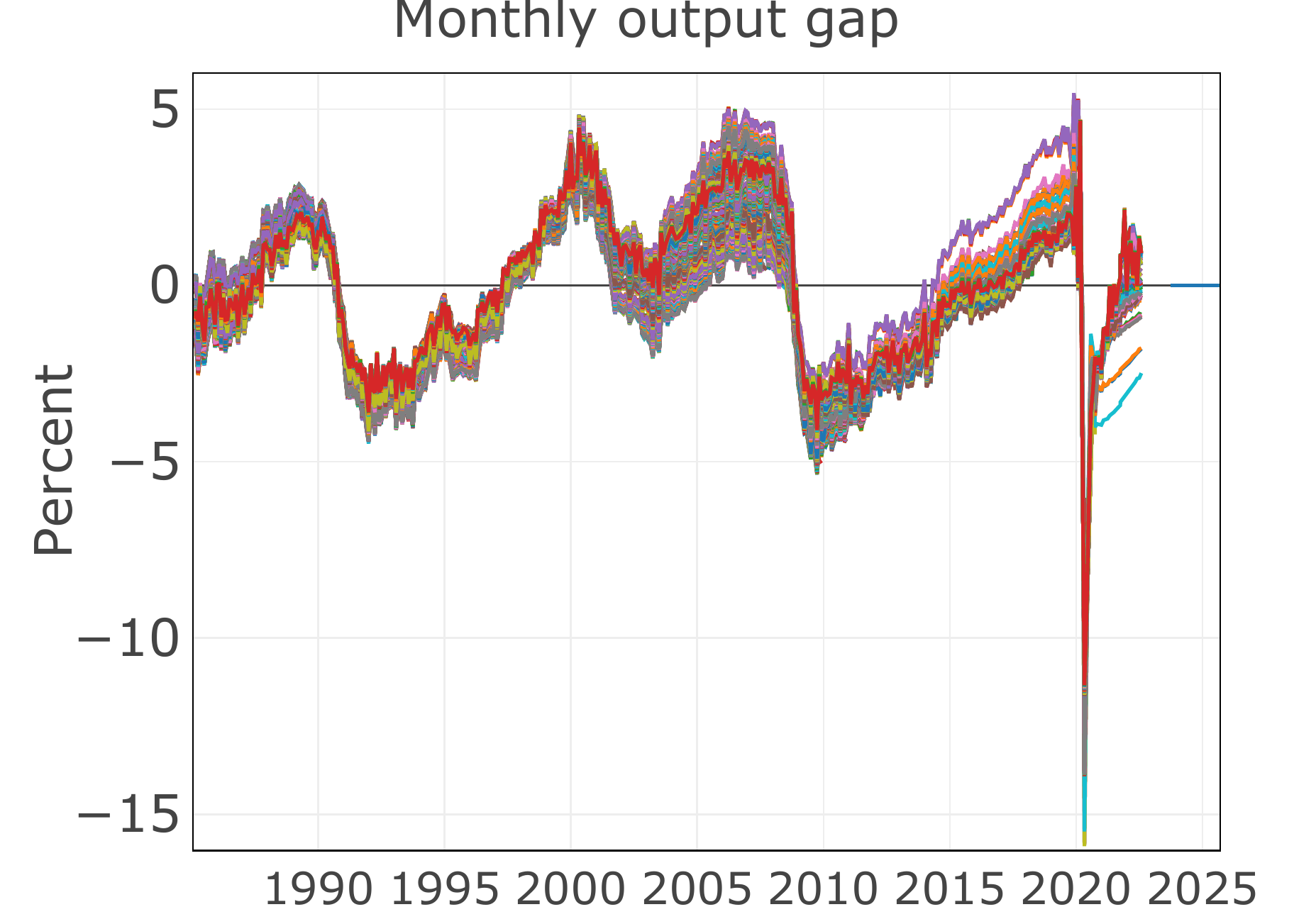}
%
	\includegraphics[width=0.49\textwidth]{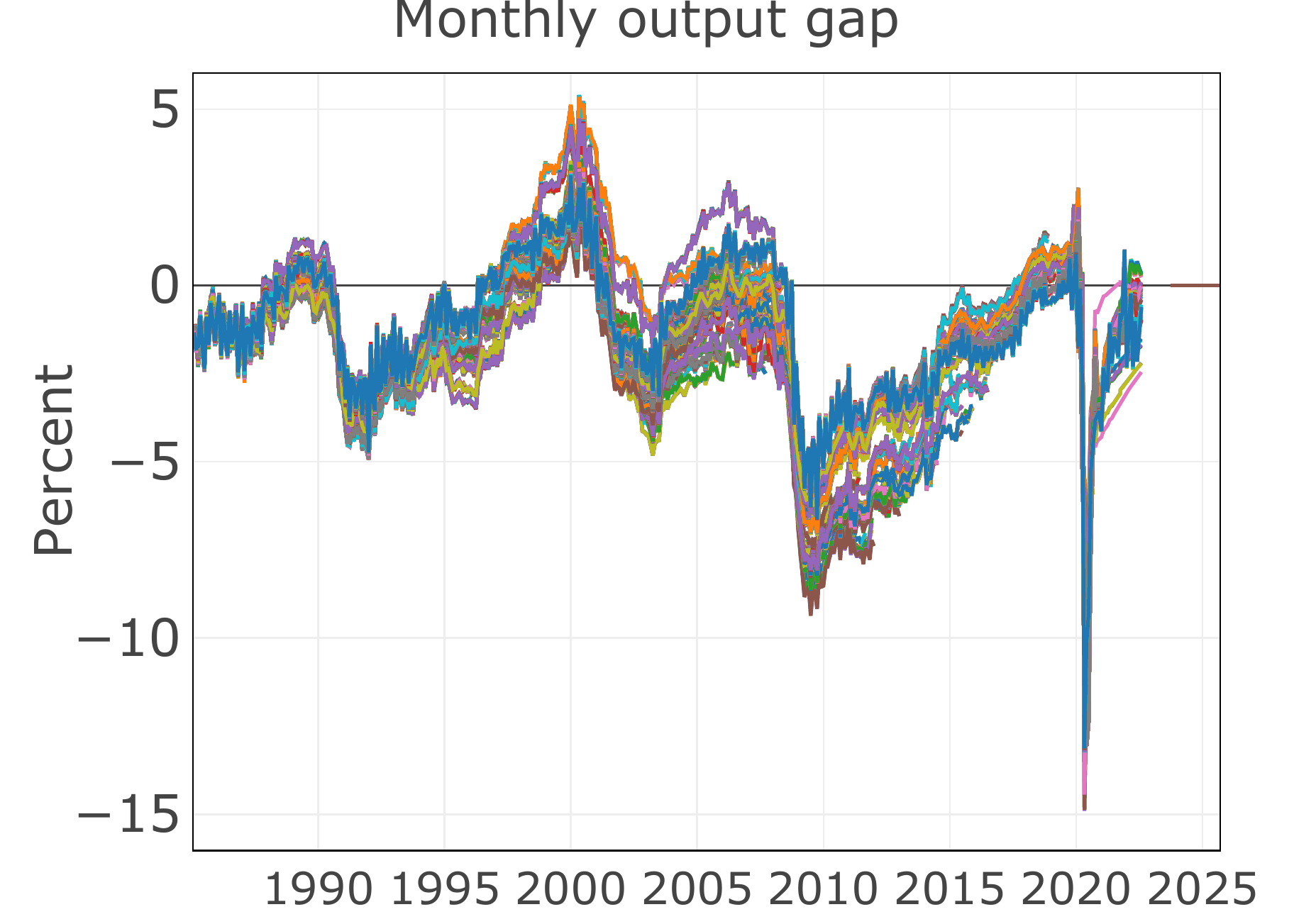}
 	\caption{The chart reports the real-time estimates of the output gap from the undisciplined (left) and the tracking model (right). The out-of-sample evaluation starts in January 2005 and ends in September 2020.}
    	\label{fig:real_time_estimates}
\end{figure}

We now turn to a real-time forecasting exercise, which is akin to a cross-validation exercise of the two models. To this end, we construct a set of real-time data vintages from \citet{alfred} and \citet{spf} starting on January 1, 2005, and using the prior 20 years as our pre-sample. We iterate over the real-time release calendar of the variables in the model and update our estimates of the trends and gaps at each new data release. We also project the trends and gaps forward and use them to forecast the variables in the model. To decrease the computational burden, we re-estimate all model parameters at the first release of each year, and then keep them fixed for the remainder of the year. 
    
The output gap estimates in \autoref{fig:real_time_estimates} are defined as the ratio of the sum of the business cycle component and the idiosyncratic cycle to the trend. For the tracking model, the sum of the business cycle and idiosyncratic components equals the output gap estimated by the CBO by construction. In the pre-sample period, before 2005, the different lines reflect the instability of in-sample estimates, while from 2005 onwards, they are also affected by data revisions. In the tracking model, revisions of the output gap can be due to revisions of the CBO estimates themselves, as reported by the CBO (or the model's revisions of the CBO gap on the forecasting horizon).

We compute two statistics to understand the relative stability of output gap measures across models. First, we calculate the standard deviation of the output gap and potential output across all vintages for each reference period. We then compute our first statistics by averaging these standard deviations across all reference periods. For the undisciplined model, this statistic is 0.54 for the output gap and 6.79 for potential output compared to 0.61 and 8.33 for the tracking model. The second statistic is the average of the maximum absolute value of revisions for each reference month. For the undisciplined model, this measure is 0.91 for the output gap and 16.38 for potential output, compared to 1.37 and 15.09 for the tracking model.

The results suggest that, on average, the estimated output gap is more stable in the undisciplined model than in the tracking model. The same is true for potential output, although the difference across models is smaller. This leads to the conclusion that the difference in variability of the gap measures across models is due to the CBO's larger judgmental revisions in the output gap and not simply the data revisions themselves. The statistics described here and those computed for the in-sample period ending at the end of 2004 are also reported in the Online Appendix, in  \autoref{tab:revisions_statistics}.

\begin{figure}[t!]
	\centering
	\includegraphics[width=\textwidth]{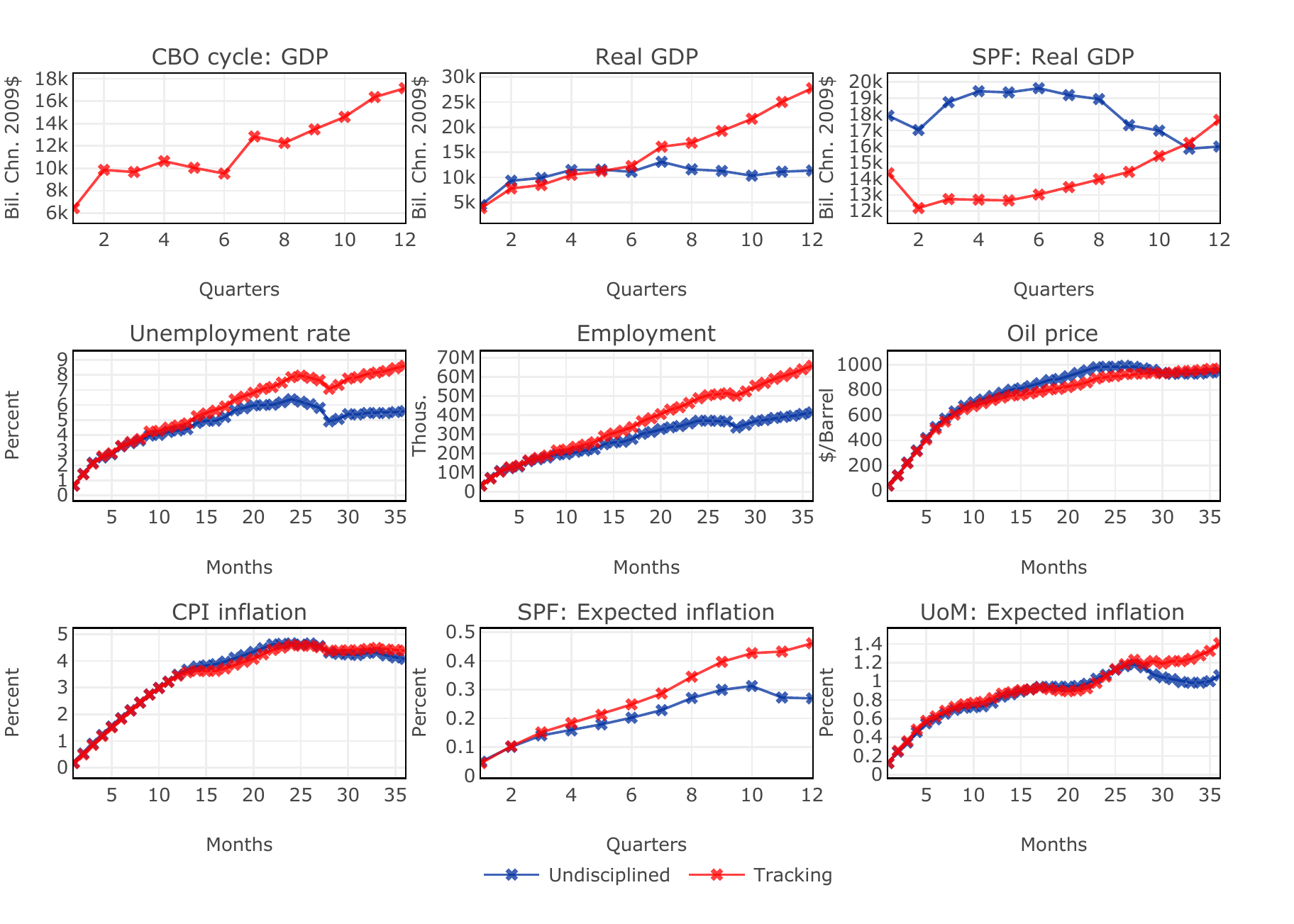}
	\caption{The chart reports the average mean squared error of the undisciplined model (in blue) and the tracking model (in red). The out-of-sample evaluation starts in January 2005 and ends in September 2020.}	
	\label{fig:mse}
\end{figure}

The COVID pandemic period also provides a good illustration of the framework's flexibility. During that period, the estimate of potential output (but for a slight uptick) tracks developments smoothly despite the enormous size of the economic shock and its unprecedented nature (see \autoref{fig:real_time_estimates}).

A comparative assessment of the out-of-sample performances of the two models is provided by the average mean squared error of forecasts up to three years ahead (see \autoref{fig:mse}). At business cycle frequency -- i.e., for horizons beyond a year --where the structural decomposition matters more, the undisciplined model outperforms the tracking model for real output, employment, unemployment rate, inflation, and inflation surveys. The tracking model only outperforms the undisciplined model up to six quarters ahead of the SPF GDP survey expectations. These results indicate that the undisciplined model is better at pinning down the business cycle components.\footnote{This may indicate that professional forecasters form their expectations by incorporating CBO estimates in their information set, but lose forecasting accuracy in doing so.} This is indeed a surprising result, considering the larger information set of the tracking model, indicating the limited value of the CBO measure for the model in the cyclical assessment of the economy.

This result sheds some light on the debate about the size of the Phillips curve since whether the latter is, on average flat or steep depends on the estimate of the size of the output gap. Based on our view of the gap, the Phillips curve is, on average, relatively steep (the slope is -0.42 in the undisciplined model and -0.36 in the tracking model). The model with the steeper curve does better at forecasting inflation. 

It is also worth observing that the Phillips curve is relatively steeper than some of the estimates in the literature for both models. This is because the models can estimate a cleaner output gap measure by separating it from a cyclical energy price component independent of local real economic conditions. The effects of energy price shocks can be seen as confounding factors in other studies, reducing the estimated correlation between the slack in the economy and price pressures.

\section{Concluding comments}

This paper proposes a framework to study the output gap in real time, via a structural medium-size time series model that is able to incorporate monthly and quarterly variables. 

We consider two specifications: a purely statistical version of the model that identifies trends and cycles via a minimum set of economically motivated restrictions; and a model that is also informed by the Congressional Budget Office's (CBO) estimate of the output gap. In the latter specification, the CBO estimates are treated as an observable variable, while multivariate correlations are estimated.

Empirical results show that the undisciplined version of the model implies, on average, a larger common component between the output gap and inflation. They also show that the undisciplined model produces more stable estimates in real-time and performs better in forecasting inflation, GDP and labor market variables beyond the very short term. The difference between the CBO's and the statistical model's estimates of the output gap are driven by a different assessment of the output potential. The statistical model identifies a decrease in the slope of output potential since 2001 and, in particular, after the financial crisis. Although the two models are ex-ante equally plausible, the facts that the model without the CBO measure does better at forecasting the medium term, and offers more stable estimates lend support to the undisciplined model. 

A by-product of our approach is to obtain a monthly version of the CBO output gap that can be updated in real time, in relation to the flow of data releases. 

The proposed framework can be easily applied to other empirical problems requiring the estimate of unobservable variables such as the natural rate of interest and can be used, as we have done for the CBO output gap, to evaluate the implications of official estimates of unobserved quantities for business cycle co-movements and forecasting. 

\clearpage
\bibliographystyle{aer}
\bibliography{bibliography}


\end{document}